\documentclass[11pt,a4paper]{article}
\usepackage{amssymb,amsfonts,amsmath,amsthm}
\usepackage{amsmath}
\usepackage[utf8]{inputenc}
\usepackage[english]{babel}
\usepackage[left=2.7cm,right=2.7cm,bottom=2.67cm,top=2.67cm]{geometry}
\usepackage{dsfont}
\usepackage{mathrsfs}
\usepackage{natbib}
\usepackage{enumerate}
\usepackage{color}
\usepackage[implicit=false]{hyperref}
\usepackage{graphicx}
\usepackage{multirow}
\usepackage{multicol}
\usepackage{subfigure}
\usepackage{placeins}
\usepackage{rotating}
\usepackage{lscape}

\newcommand{\E}{\mathds{E}}
\newcommand{\F}{\mathscr{F}}
\newcommand{\R}{\mathds{R}}
\newcommand{\N}{\mathds{N}}
\newcommand{\Z}{\mathds{Z}}
\newcommand{\eps}{\varepsilon}
\newcommand{\bs}[1]{\boldsymbol{#1}}

\long\def\sfootnote[#1]#2{\begingroup%
\def\thefootnote{\fnsymbol{footnote}}\footnote[#1]{#2}\endgroup}

\def\bfootnote{\xdef\@thefnmark{}\@footnotetext}

\begin{document}
\pagestyle{myheadings} % allow the use of headings
\markboth{Order selection in GARMA models}{K.Z. Lastra. G. Pumi and T.S. Prass}

\thispagestyle{empty}
{\centering
\Large{\bf Order selection in GARMA models for count time series: a Bayesian perspective}\vspace{.5cm}\\
\normalsize{ {\bf Katerine Zuniga Lastra${}^\mathrm{a}$,\let\thefootnote\relax\footnote{\hskip-.3cm$\phantom{s}^\mathrm{a}$Mathematics and Statistics Institute and Programa de P\'os-Gradua\c c\~ao em Estat\'istica - Federal University of Rio Grande do Sul.
%- 9500,  Bento Gon\c calves Avenue - 91509-900, Porto Alegre - RS - Brazil.
} Guilherme Pumi${}^{\mathrm{a,}}$\sfootnote[1]{Corresponding author. This Version: \today} and Taiane Schaedler Prass${}^\mathrm{a}$
 \\
\let\thefootnote\relax\footnote{E-mails: zunigalastrakaterine@gmail.com (K.Z. Lastra), guilherme.pumi@ufrgs.br (G. Pumi), taiane.prass@ufrgs.br (T.S. Prass)}}\\
\vskip.3cm
}}
% \affiliation{Universidade Federal do Rio Grande do Sul}
% \address[a]{Department of Statistics and Graduate Program in Statistics \\
% Universidade Federal do Rio Grande do Sul\\
% 91509-900, Porto Alegre\\
% Brazil\\
% \printead{e1,e2}}

%%%%%%%%%%%%%%%%%%%%%%%%%%%%%%%
%%  \Bigg(\bigg(\Big(\big(   %%
%%%%%%%%%%%%%%%%%%%%%%%%%%%%%%%

\begin{abstract}
Estimation in GARMA models has traditionally been carried out under the frequentist approach. To date, Bayesian approaches for such estimation have been relatively limited. In the context of GARMA models for count time series, Bayesian estimation achieves satisfactory results in terms of point estimation. Model selection in this context often relies on the use of information criteria. Despite its prominence in the literature, the use of information criteria for model selection in GARMA models for count time series have been shown to present poor performance in simulations, especially in terms of their ability to correctly identify models, even under large sample sizes. In this work, we study the problem of order selection in GARMA models for count time series, adopting a Bayesian perspective considering the Reversible Jump Markov Chain Monte Carlo approach. Monte Carlo simulation studies are conducted to assess the finite sample performance of the developed ideas, including point and interval inference, sensitivity analysis, effects of burn-in and thinning, as well as the choice of related priors and hyperparameters. Two real-data applications are presented, one considering automobile production in Brazil and the other considering bus exportation in Brazil before and after the COVID-19
pandemic, showcasing the method's capabilities and further exploring its flexibility.
\vspace{.2cm}

\noindent \textbf{Keywords:} Count time series; Regression models; Bayesian analysis; Reversible Jump Markov Chain.\vspace{.2cm}\\
\noindent \textbf{MSC:} 62M10, 62F15, 62J02, 62F10.
\end{abstract}
\section{Introduction}
Counting time series typically arise when the interest lies in the count of certain events happening during times intervals. They are ubiquitous to all fields of study and applications abundant. For instance, \cite{Zeger} and \cite{davis2000autocorrelation} analyzed the incidence of certain diseases. In the field of insurance,  \cite{freeland2004analysis} presented an application to the monthly count data set of claimants for wage loss benefit, in order to estimate the expected duration of claimants in the system. \cite{liesenfeld2006modelling} studied fluctuations in the financial market whereas \cite{weiss2007controlling} considered time series of count in the context of quality management strategies and  \cite{brannas1994time} modeled the number of traffic accidents in a given location.

Models for time series of count are mainly modeled under the frameworks of parameter and observation driven models, according to Cox's classification \cite{cox1981}. The former extends generalized linear models by incorporating a latent  process into the conditional mean of the counting process, while the latter directly rely on the count observed in each interval to discern the temporal dynamics, specifying a model for the distribution of the count at each moment.

Among the modeling approaches, the class of GARMA (generalized autoregressive moving average) models, introduced under this name by \cite{benjamin2003}, have been extensively studied in the recent literature, being considered one of the most promising approaches to non-Gaussian time series modeling. GARMA are observation driven models which merge the classical ARMA modeling approach within the flexibility of the Generalized Linear Model (GLM) framework. Inference in GARMA models are usually conducted under a frequentist framework, based on conditional or partial likelihood. The literature considering Bayesian inference in GARMA models is less abundant. For instance, in the context of continuously distributed GARMA models, \cite{casarin} and \cite{aline2} consider Bayesian inference in the context of the $\beta$ARMA of \cite{rocha2009}. \cite{bgarma} considers Bayesian inference in the context of GARMA models for count time series in the classical framework, namely, when the conditional distribution is a member of the canonical exponential family. {\cite{PCS} is closely related to \cite{bgarma}, but considering the negative binomial with both parameters unknown and also studying the Poisson inverse gaussian GARMA model, whereas \cite{TGARMA} considers Bayesian estimation in the context of transformed GARMA models. \cite{ZM} considers Bayesian estimation of Zero-Modified Power Series GARMA in the context of count time series that exhibit zero inflation or deflation. \cite{BMC} considers a Bayesian Poisson and Negative Binomial GARMA as candidates to model temporal random effects in a spatial temporal analysis of infant mortality in Africa. }

One important matter is model selection for GARMA models. In the frequentist framework this is usually attained by using either a Box and Jenkings-like approach or by using information criteria. information criteria are also widely applied in the context of Bayesian model selection. One important alternative is the so-called  Reversible Jump Markov Chain Monte Carlo (RJMCMC) approach, introduced by \cite{green}. The RJMCMC is an extension of the Metropolis-Hastings algorithm allowing the generation of samples from a target distribution in spaces of different dimensions. To the best of our knowledge, the only work considering an RJMCMC approach in the context of GARMA models is \cite{casarin} which considered model selection using an RJMCMC approach in the context of a subclass of $\beta$AR models.

In this paper, we propose and discuss model selection in GARMA models for count time series using an RJMCMC approach. The commonly applied general purpose RJMCMC can be adapted to be used in the context of GARMA models following the approach employed by \cite{troughton} in the context of ARMA$(p,q)$ models. The main idea is to enumerate the possible combinations of model orders and use this enumeration as index for model transition. We shall consider a different approach however, in which transitions are determine by inclusion/exclusion of each parameter, given the current state of the chain, according to a prior inclusion probability. The proposed approach allow for more flexibility in model configuration, widening the scope of possible models to be visited by the chain.

The paper is organized as follows. In Section 2, a review of the GARMA model class is conducted, addressing key concepts related to Bayesian inference for these models and the RJMCMC method. In Section 3, we carry out a Monte Carlo simulation to assess the finite sample performance of the proposed Bayesian approach, with emphasis on model selection. In section 4 we present two real data applications of the proposed methodology. Lastly, we present our conclusions.
\section{GARMA Models}
Let $\{Y_t\}_{t\in\Z}$ be a stochastic process of interest and let $\{\bs x_t\}_{t\in\Z}$ be a set of $r$-dimensional exogenous covariates to be included in the model. Let $\F_{t}=\sigma\left\{\bs x_{t}', \bs x_{t-1}', \cdots, Y_{t-1}, Y_{t-2}, \cdots \right\}$ be the information available to the observer at time $t$. In \cite{benjamin2003}, GARMA models are considering that the distribution of $Y_t$ given the information observed up to time $t$ belongs to the exponential family in canonical form, that is
\begin{equation} \label{Garma}
    f\big(y;\omega_t,\varphi|\F_{t-1}\big)=\exp\biggl\{\frac{y\omega_{t}-b(\omega_t)}{\varphi}+c(y,\varphi)\biggr\},
\end{equation}
where, $\omega_{t}$ and $\varphi$ are the canonical and scale  parameters, respectively, with $b(\cdot)$ and $c(\cdot)$ being specific functions that define the particular exponential family. In traditional GARMA models, $\omega_t$ is time dependent while $\varphi$ is not, which is reflected in the notation. The conditional mean and variance of $Y_{t}$, given $\F_{t-1}$, are given by $\mu_{t} = E\big(Y_{t} | \F_{t-1}\big)=b^{\prime}\big(\omega_{t}\big)$ and $\operatorname{Var}\big(Y_{t} | \F_{t-1}\big)=\varphi b^{\prime \prime}\big(\omega_{t}\big)=\varphi V\big(\mu_{t}\big)$, with $t \in\{ 1,\cdots,n\}$. In the systematic component of the model, the conditional mean $\mu_{t}$ is related to the linear predictor possibly through a twice differentiable invertible link function $g$. The most commonly used structure for the systematic component  includes covariates and an ARMA structure of the form
\begin{equation*} %\label{PL}
    \eta_t = g(\mu_{t}\big)=\alpha+\bs x_{t}^{\prime} \bs\beta+\sum_{j=1}^{p} \phi_{j}\big[g\big(Y_{t-j}\big)-\bs x_{t-j}^{\prime} \bs \beta\big]+\sum_{j=1}^{q} \theta_{j}\big[g\big(Y_{t-j}\big)-\eta_{t-j}\big],
\end{equation*}
where $\eta_t$ is the linear predictor, $\alpha$ is an intercept, $\bs\beta=(\beta_1, \cdots,\beta_r)^\prime$ is the parameter vector related to the covariates, $\bs\phi=(\phi_1,\cdots,\phi_p)^\prime$ and $\bs\theta=(\theta_1,\cdots,\theta_q)^\prime$ are the AR and MA coefficients, respectively. A GARMA$(p,q)$ model is defined by \eqref{Garma} and \eqref{PL}.

The component \eqref{Garma} can be continuous \citep[][among others]{rocha2009,Bayers,helen}, discrete \citep[][among others]{benjamin2003,airlane,airlane2}, or even of the mixed type \citep{ibarma}. The most commonly applied GARMA models for time series of counts are reviewed in the next section.

%%%%%%%%%%%%%%%%%%%%%%%%%%%%%%%%%%%%%%%%%%%%%%%%%%%%%%%%%%%%%%%%%%%%%%%%%%%%%%%%%%%%%%%%%%%%%%
\subsection{GARMA models for time series of counts}
In this Section, we present a brief description of the three most applied GARMA models for counting data, namely, the Poisson GARMA, binomial GARMA, and negative binomial GARMA models. {These apply the the logarithm as link function, which require a small adaptation to avoid numerical instability, namely,
\begin{equation}\label{PL}
    \log (\mu_{t})=\alpha + \bs x_{t}^{\prime} \bs\beta+\sum_{j=1}^{p} \phi_{j}\big[\log (Y_{t-j}^{*})-\bs x_{t-j}^{\prime} \bs\beta\big]+\sum_{j=1}^{q} \theta_{j}\big[\log(Y^{*}_{t-j})-\log(\mu_{t-j})\big],
\end{equation}
where $Y^{*}_{t}=\max (Y_{t}, c)$, for $0<c<1$, is a user-defined threshold applied to avoid numerical problems. The conditional distribution and \eqref{PL} define each particular GARMA model.}
%==================================================================================

\subsubsection{Poisson GARMA model}

When $Y_{t}|\F_{t-1}$ follows a Poisson distribution with
mean $\mu_t$ we have
{
\begin{equation} \label{PModel1}
    f\big(y_t;\mu_t|\F_{t-1}\big)=\exp\big\{y_t \log (\mu_{t})-\mu_{t}-\log (y_t!)\big\}I(y_t\in\N),
\end{equation}}
which belongs to the canonical exponential family with
\begin{equation*}
{\varphi=1, \quad \omega_{t}=\log (\mu_{t}), \quad b(\omega_{t})=e^{\omega_t}, \quad c(y_{t}, \varphi)=-\log (y_{t} !), \quad \mu_{t}=e^{\omega_{t}}, \quad \text{and}\ \ V(\mu_{t})=\mu_{t}.}
\end{equation*}
%==================================================================================
\subsubsection{Binomial GARMA model}
When $Y_{t}|\F_{t-1} \sim B\big(m, p_{t}\big)$, with $m>0$ known, and $\mu_t=\E(Y_{t}|\F_{t-1})=mp_t$ we have
\begin{equation}\label{BModel1}{
    f\big(y_t;\mu_t|\F_{t-1}\big)=\exp \biggl\{y_t \log \biggl(\frac{p_{t}}{1-p_{t}}\biggr)+m \log(1-p_{t})+\log \biggl(\frac{\Gamma(m+1)}{\Gamma\big(y_t+1\big) \Gamma\big(m-y_t+1\big)}\biggr)\biggr\},}
\end{equation}
which is a member of the canonical exponential family with $\varphi=1$,
\begin{align*}
&{\omega_{t}=\log \bigg(\frac{\mu_{t}}{m-\mu_{t}}\bigg),\quad  b\big(\omega_{t}\big)=m \log \bigg(\frac m{m-\mu_{t}}\bigg),\quad
c\big(y_{t}, \varphi\big)=\log \biggl(\frac{\Gamma(m+1)}{\Gamma\big(y_t+1\big) \Gamma\big(m-y_t+1\big)}\biggr),}\\
&{\mu_{t}=\frac{m\exp(p_{t})}{1+\exp(p_{t})} \quad \text{and} \quad  V(\mu_{t})=\frac{\mu_{t}(m-\mu_{t})}{m}.}
\end{align*}
%The binomial GARMA model is specified by \eqref{BModel1} and \eqref{PL}.
%==================================================================================
\subsubsection{Negative binomial GARMA model}
When $Y_{t}|\F_{t-1} \sim\mathrm{NB}\big(k, p_{t}\big)$, with $k>0$ known, we have $\E(Y_{t}|\F_{t-1}) =  \frac{k(1-p_t)}{p_t}$, so that $p_t=\frac k{\mu_t+k}$ and hence
\begin{equation} \label{BNModel1}{
    f\big(y_t;\mu_t|\F_{t-1}\big)=\exp \Bigg\{k \log \bigg(\frac{k}{\mu_{t}+k}\bigg)+y_t \log \bigg(\frac{\mu_{t}}{\mu_{t}+k}\bigg)+\log \biggl(\frac{\Gamma(k+y_t)}{\Gamma(y_t+1) \Gamma(k)}\biggr)\biggr\},}
\end{equation}
which belongs to the exponential family with
\begin{align*}
&{\varphi=1,\quad \omega_{t}=\log \bigg(\frac{\mu_{t}}{\mu_{t}+k}\bigg),\quad b\big(\omega_{t}\big)=-k \log \biggl(\frac{k}{\mu_{t}+k}\biggr),\quad c\big(y_{t}, \varphi\big)=\log \biggl(\frac{\Gamma(k+y_t)}{\Gamma(y_y+1) \Gamma(k)}\bigg),}\\
&{\mu_{t}=\frac{k\exp(p_{t})}{1+\exp(p_{t})} \quad \text{and} \quad V(\mu_{t})=\frac{\mu_{t}(k+\mu_{t})}{k}.}
\end{align*}
\subsubsection{Bayesian approach to GARMA modeling}
The partial likelihood function for the model is given by
\begin{equation}\label{LL}
    \mathcal{L}\big(\bs\phi, \bs\theta, \alpha_{0} |  \F_{t}\big) \propto \exp \bigg\{\sum_{t=s+1}^{n} \frac{y_{t} \theta_{t}-b\big(\omega_{t}\big)}{\varphi}+c\big(y_{t}, \varphi\big)\bigg\},
\end{equation}
where $\omega_{t}$ is the canonical parameter of the model and $s$ is the starting point of the likelihood function, most often taken as $s=0$ as in \cite{benjamin2003} and \cite{PUMI2019} but sometimes taken as $s=\max\{p,q\}$ as in \cite{rocha2009} and \cite{bgarma}. In this work we shall employ $s=0$.

For $\alpha_0$, $\bs\phi$ and $\bs\theta$, we shall assume normal prior distributions with zero mean and variance $\sigma^{2}$ for each component, that is $\phi_{i} \sim N\big(0, \sigma^{2}\big)$,  $\theta_{j} \sim N(0, \sigma^{2})$ and $\alpha_{0} \sim N(0, \sigma^{2})$, for $i\in\{1, \cdots, p\}$ and $j=\{1, \cdots, q\}$. Assuming independence between the parameters, the joint prior distribution is
\begin{equation} \label{aPrioriBayGM}
\pi_{0}\big(\bs\phi, \bs\theta, \alpha_{0}\big) \propto\exp\bigg\{-\frac{1}{2 \sigma^{2}}\bigg(\alpha_{0}^{2}+\sum_{i=1}^{p} \phi_{i}^{2}+\sum_{j=1}^{q} \theta_{j}^{2}\bigg)\bigg\}.
\end{equation}
Therefore, the posterior conditional distribution for the model is written as
\begin{equation*}
    \pi\big(\bs\phi, \bs\theta, \alpha_{0} |\F_{t}\big) \propto \mathcal{L}\big(\bs\phi, \bs\theta, \alpha_{0} | \F_{t}\big) \pi_{0}\big(\bs\phi, \bs\theta, \alpha_{0} \big).
\end{equation*}
{Explicit formulae for the posterior distribution are straightforwardly obtained form the likelihood functions of each proposed model, which, in turn, are derived by substituting the conditional densities from equations \eqref{PModel1}, \eqref{BModel1} and \eqref{BNModel1} into \eqref{LL}. }
\subsubsection{Reversible-jump Markov chain Monte Carlo}
As mentioned in the introduction, the method known as Reversible-jump Markov chain Monte Carlo (RJMCMC), introduced by \cite{green},  is an extension of the Metropolis-Hastings algorithm allowing the generation of samples of a target distribution in spaces of different dimensions. The dimension of the parameter space is allowed to vary between iterations and is commonly used as a Bayesian method for model selection.
According to \cite{green} in a Bayesian modeling context, one has a countable collection of candidate models $\left\{M_k, k \in K\right\}$, where the index $k$ serves as an auxiliary indicator variable of the model and $K$ represent the scope of the considered models. The model $M_k$ has a vector of $k+1$  unknown parameters, say $\bs\xi_k \in R^{k+1}$, that can assume different values for different models. There is a natural hierarchical structure expressed by modeling the joint distribution of $(k, \bs\xi_k, y)$ as
\begin{equation*}
   p(k, \bs\xi_k, y) \propto p(k) p(\xi_k | k) p(y|\bs\xi_k, k).
\end{equation*}
The Bayesian inference about $k$ and $\bs\xi_k$ will be based on the posterior distribution $p(k, \bs\xi_k|y)$, given by
\begin{equation*}
    p(\bs\xi_k|y, k) \propto p(y|\bs\xi_k, k) p(\bs\xi_k|k)
\end{equation*}
Where $p(y|\bs\xi_k, k)$ and $p(\bs\xi_k|k)$ represent the probability model and the prior distribution of the model parameters $M_k$, respectively. Thus, the posterior probability is given as,
\begin{equation*}
    p(k, \bs\xi_k|y) \propto p(k) p(\bs\xi_k|k, y)
\end{equation*}
According to \cite{casarin}, the posterior joint distribution is the target distribution of the RJMCMC sampler over the state space $\Theta=\cup_{k \in K}\big(k, {\R}^{n_k}\big)$. Within each iteration, the RJMCMC algorithm updates the parameters given the model order and then the model order given the parameters. If the current state of the Markov chain is $(k, \bs\xi_k)$, then a possible version of the RJMCMC algorithm is as follows:
%==================================================================================
\subsection*{General RJMCMC algorithm}
\begin{itemize}
\item[]Step 1. Propose a visit to model $M_{k'}$ with probability $J\left(k \rightarrow k^{\prime}\right)$.
\item[]Step 2. $\nu$ is sampled from a proposal density $q\left(\nu | \bs\xi_k, k, k^{\prime}\right)$.
\item[]Step 3. Set $\left(\bs\xi_{k^{\prime}} ,\nu^{\prime}\right)=g_{k, k^{\prime}}\left(\bs\xi_{k^{\prime}}, \nu\right)$, where $g_{k, k^{\prime}}(\cdot)$ is a bijection between $\left(\bs\xi_{k}, \nu\right)$ e $\left(\bs\xi_{k^{\prime}}, \nu^{\prime}\right)$.
\item[]Step 4. The acceptance probability of the new model is
\begin{equation*}
    \alpha_{k \rightarrow k^{\prime}}=\min \bigg\{1, \frac{p(y|k^{\prime}, \bs\xi_{k'}) p(\bs\xi_{k'}) p(k^{\prime}) J(k^{\prime} \rightarrow k) q(\nu^{\prime}|\bs\xi_{k^{\prime}}, k^{\prime}, k)}{p(y|k, \bs\xi_k) p(\bs\xi_k) p(k) J(k \rightarrow k^{\prime}) q(\nu|\bs\xi_{k}, k^{\prime}, k)} \times\bigg|\frac{\partial g_{k, k^{\prime}}(\bs\xi_k, \nu)}{\partial(\bs\xi_{k^{\prime}}, \nu)}\bigg|\bigg\}.
\end{equation*}
\end{itemize}
Looping through steps 1–4 generates a sample $\{k_l, l=1, \cdots, L\}$  for the model indicators and
\begin{equation*}
    \hat{p}(k|y)=\frac{ 1}{L} \sum_{l=1}^L I(k_l=k)
\end{equation*}
where $I(\cdot)$ is the indicator function. In practice, $J\left(k \rightarrow k^{\prime}\right)$ is usually taken as $N(0, \sigma^2)$, where $\sigma^2$ is a scale hyperparameter.

In the case of ARMA models, the general RJMCMC algorithm can be applied by indexing the model order $(p,q)$ by means of a bijection between the scope of models of interest, say $\{(p,q)\in\N^2:1\leq p\leq p_m, 1\leq q\leq q_m\}$ for $p_m$ and $q_m$ the maximum values of $p$ and $q$ desired, and the positive integers. In this way, the RJMCMC algorithm for ARMA follow essentially steps 1 through 4 above, as presented in \cite{troughton} and extended to ARFIMA models by \cite{eugri}. The same approach can, in principle, be used in the context of GARMA models. One criticism to this approach is that transitioning between models via the indexing of $(p,q)$ implies that only ``complete'' models are considered in each transition. This constraint may be somewhat limiting, especially when exploring the entire scope of possible ARMA submodels. Additionally, the implementation of this approach can be challenging and less generalizable due to the need for careful indexing.

Instead, we propose a more direct and simplified approach to the RJMCMC for GARMA models. This method not only facilitates a more thorough exploration of GARMA submodels but is also easier to implement using widely available general RJMCMC packages and software. We start by determining values $p_m$ and $q_m$ for which the the most complex model of interest is of order $(p_m,q_m)$. Let $\bs\phi_m:=(\phi_1,\cdots,\phi_{p_m})'$ and $\bs\theta_m:=(\theta_1,\cdots,\theta_{q_m})'$ be the associated AR and MA parameters, respectively. Transitions from one model to another occur by determining whether each parameter $\phi_i$, for $i\in\{1,\cdots,p_m\}$, is to be included in the model or not, according to a prior inclusion probability, given the current chain state. If a particular $\phi_i$ is to be included in the model, then it is sampled normally. Otherwise the parameter is set to 0. The procedure is repeated to cover parameters $\theta_j$, $j\in\{1,\cdots,q_m\}$.

By proposing, transition by transition, which parameters to include in the model (given the current state), the algorithm explores model configurations that are rarely considered in practice. For instance, for $p_m=q_m=3$, the algorithm might propose a model for which only $\phi_3$ and $\theta_3$ are different from 0. The selection of $p_m$ and $q_m$ are important in this context, given the potential for $2^{p_m+q_m}$ submodels that can be proposed using this approach. While larger values of $p_m$ and $q_m$ may increase the algorithm's flexibility, they also present a challenge as the resulting scope of possible models may be too extensive, requiring very large chains for the MCMC sampler to converge.
\section{Simulations}
In this section, we present a Monte Carlo simulation study aimed at evaluating the finite sample performance of the proposed model selection in the context of GARMA$(p,q)$ count models. In the simulation, we consider the point and interval estimation of the parameters of interest and also the percentage of models correctly selected by the proposed approach.

As expected, given the characteristics of the RJMCMC, and widely reported in the literature, samples from the posterior distribution  obtained via RJMCMC are typically sensible to the initial values and to the scale hyperparameter $\sigma^2$, associated with the transition probability and are highly correlated as well {\citep{green,RichardGreen,brooks1998,HastieGreen,Dellaportas,Gelman2013}}. Mitigating the influence of initial values in the posterior sample is usually attained through a burn-in, whereas, autocorrelation in the sample can be mitigated by using a thinning approach. With that in view, we also provide a sensitivity analysis with respect to the burn-in, thinning and the scale hyperparameter used. The simulation was carried out using the software \texttt{R} \citep{R}, version 4.0.3. To perform the RJMCMC, we use \texttt{R} package \texttt{Nimble} \citep{nimble}. {For first time users of \texttt{Nimble}, there are detailed and comprehensive online information regarding the package's use, including examples. We recommend the project's Github {\color{blue}\href{https://github.com/nimble-dev/nimble}{github.com/nimble-dev/nimble} and the dedicated webpage {\color{blue}\href{https://r-nimble.org/}{r-nimble.org}}}. Globally, the implementation of our model follows the usual use of \texttt{Nimble}'s RJMCMC module. One exception is in function \texttt{configureMCMC}, where we set the boolean \texttt{useConjugacy} to false.  Any other specific non-default or user chosen value used in our implementation will be provided in the text.}
\subsection{Effects of Burn-in} \label{BI}
In this section, we examine the finite sample performance of point and interval estimation of the proposed Bayesian approach for the GARMA Binomial model with different values of $p$ and $q$, different values of the hyperparameter $\sigma\in\{0.5,5,10,15\}$ and burn-in values $\{0,1000,3000,5000\}$. Observe that the proposed RJMCMC approach perform model estimation and point estimation at the same time, hence, being different than the Bayesian approach presented in \cite{bgarma}, {where the authors fit a model and, afterwards, perform model selection, based on information criteria}.
\subsubsection{GAR$(p)$ models} \label{garp}
The first set of experiments considers GAR$(1)$ models with $(\alpha, \phi)= (-0.5, -0.4)$ and GAR$(2)$ models with $(\alpha, \phi_1,\phi_2) =(-1, 0, -0.4)$ and $m=15$. To generate the time series, a burn-in of 100 points was considered and a constant of $c=0.3$ for the binomial GARMA models was used, independently of the model considered. We generated time series of size $n=1{,}000$ and a total of 1,000 replications of each scenario were performed.

In all scenarios, RJMCMC was performed considering maximum orders $p_m=3$ and $q_m=3$ with a non-informative prior probability of $0.5$ for the inclusion of each parameter. We consider a $N(0,0.3^2)$ prior for $\alpha$ and a $N(0,0.2^2)$ for the AR parameters. These can be considered somewhat informative, but larger values of the hyperparameters were found to cause numerical instability when compiling the \texttt{Nimble} code, often making compilation impossible. We consider zero-mean normally distributed reversible jump proposals with standard deviation {(scale) $\sigma\in\{0.5, 5, 10,15\}$.} For each scenario, a single chain containing 30,000 iterations was generated.

Credible intervals (CIs) were obtained using two methods: the highest posterior density interval (HPD),  {and the empirical credible interval (ECI), based on the sample from the posterior distribution obtained. The HDP interval contains the most probable values of the posterior distribution, and it is defined as the region of the posterior distribution where the density is higher than outside this region, and it includes the specified proportion of the posterior probability (1 minus the confidence level). On the other hand, the ECI is computed based on quantiles of the posterior sample and typically represents the central region of the posterior distribution.} To obtain HPD intervals, we use function \texttt{emp.hpd} from the \texttt{R} package \texttt{TeachingDemos} \citep{TD}, while for empirical credible intervals, we apply the \texttt{R} function \texttt{quantile}. All credible intervals are presented considering a confidence level of 0.05. To calculate the effective sample size (ESS) for each parameter, we use function \texttt{effectiveSize} from the \texttt{R} package \texttt{coda} \citep{coda}.

{The simulation results are presented in Figure \ref{fig:table1} and \ref{fig:table2} below and on Tables \ref{burn-inGAR1} and \ref{burn-inGAR2} in the Appendix. The plots present the boxplots of the posterior distribution's mean along the 1,000 replications for each value of $\sigma$ (columns), parameter (rows) and burn-in size (cell). The blue lines represent the simulated (true) parameter. Tables \ref{burn-inGAR1} and \ref{burn-inGAR2} present point estimates obtained as the average (Mean), median (Median), and standard deviation (sd) of the posterior distribution, along with the average HPD intervals, obtained by averaging the limits of the credible intervals. Since the targeted posterior distribution is unimodal, the average HPD reflects the region of highest density around the parameter's true value, averaged over the replications. It is useful as a summary measure of the credible interval across replications.}

\begin{figure}
    \centering
    \includegraphics[width=\textwidth]{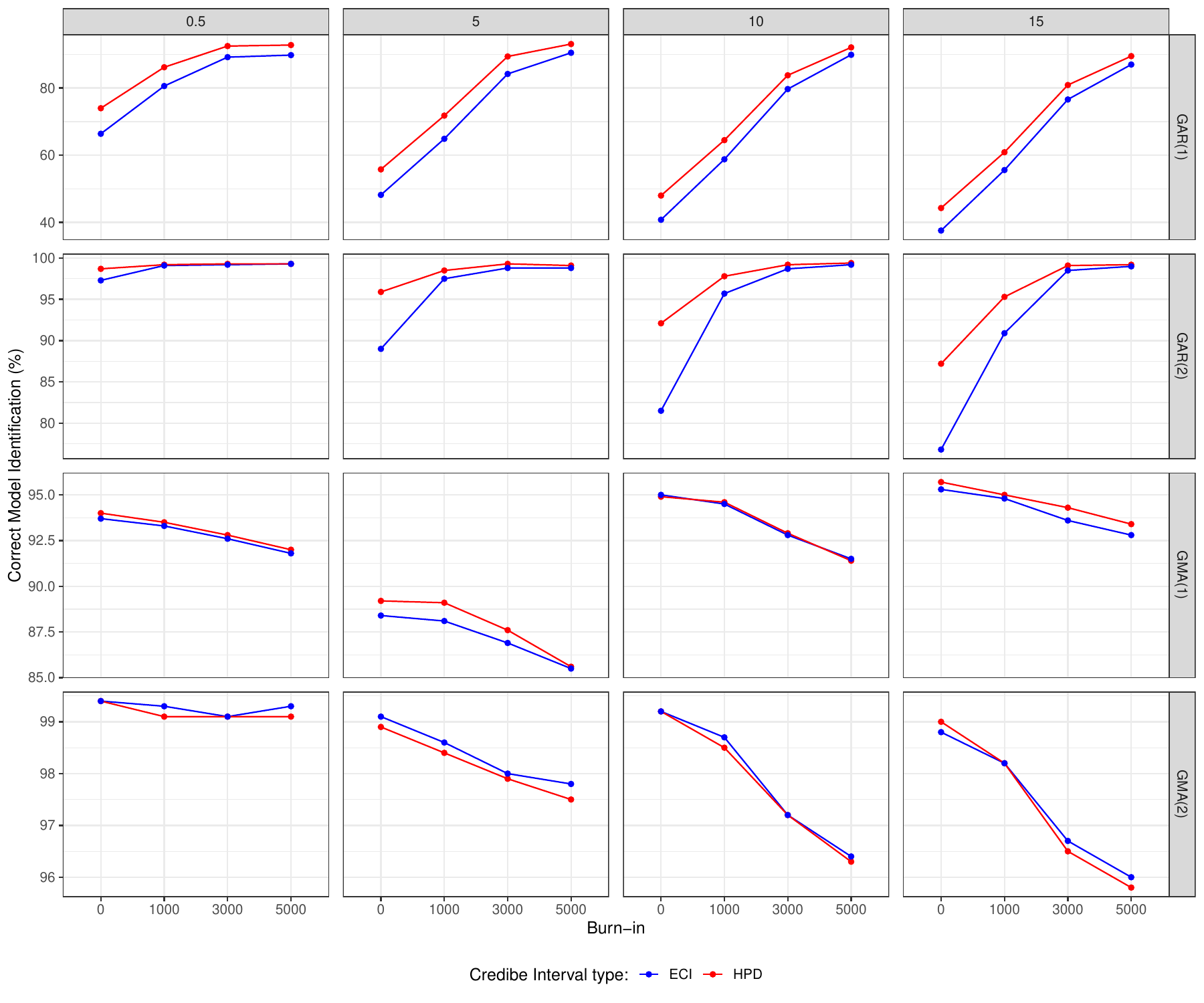}
    \caption{Correct model identification (\%) of each type of credible interval as a function of burn-in, grouped by model (rows) and $\sigma$ (columns).}
    \label{fig:cov_bi}
\end{figure}

For each parameter, we included in the tables the frequency for which the CIs correctly identify the {model} according to the data generating process, {whereas Figure \ref{fig:cov_bi} presents the percentage of correct model identification of each type of credible interval as a function of burn-in, grouped by model and $\sigma$. In the plots and tables, a value of 99\% indicates that in 990 out of 1,000 replications, the model was correctly identified by the CI. Specifically, this means that the non-zero parameters were identified as non-zero, and the non-significant parameters were correctly identified as non-significant by the CI.}
{When no burn-in is applied, from Table \ref{burn-inGAR1} and Figure \ref{fig:table1}, we observe that as the value of $\sigma$ increases, so does the bias in the estimates for the GAR$(1)$ model. For the GAR$(2)$ (Table \ref{burn-inGAR1} and Figure \ref{fig:table1}),} a pattern is not so easily identifiable and the effect of $\sigma$ in the estimation is less noticeable. In this case, the smallest bias for the non-zero parameters was obtained for $\sigma=10$. From the tables, little difference is observed when we apply the mean or median to obtain point estimation, with a slight advantage for the median in both cases. Effective sample size is low for most parameters, especially for the non-zero ones, due to high correlation in the sample. For the GAR$(1)$, effective sample size seems to increase as $\sigma$ decreases. The percentage of correctly identified models is lower for the GAR$(1)$ model than for the GAR$(2)$ and this percentage seems to decrease as $\sigma$ increases in both cases. From {(Figure \ref{fig:cov_bi}) (first and second row)}, when the HPD credible intervals are considered for model identification, a higher percentage of correctly identified models is obtained when compared to the quantile based CIs (ECI).  The best scenario in this metric was when $\sigma=0.5$ with an advantage of almost 30\% to the worst case for the GAR$(1)$ case for both CIs. For the GAR$(2)$ model these numbers are about 10\% considering HPD and about 20\% for the quantile based CIs. For the GAR$(2)$ models, the percentage of correctly identified models is fairly high, above 97\% for both CIs, but for the GAR$(1)$ it can be considered on the low end. As for standard deviations, these do not seem to be impacted by $\sigma$.

Applying a burn-in improves the results in all cases and in all metrics. The most interesting feature, however, is that the effect of $\sigma$ is highly mitigated upon applying a burn-in, yielding more dependable results overall. This is especially observed in the percentage of correctly identified models, which in the case of GAR$(1)$  increases from fairly low values to values around 90\% in all cases {(Figure \ref{fig:cov_bi})}. For the GAR$(2)$ these values are around 99\% in all cases. Effective sample size also generally increases upon applying a burn-in, but in most cases the improvement is marginal.

Regarding the size of the burn-in, for the GAR$(1)$ the improvements obtained from applying a size 3{,}000 burn-in compared to 1{,}000 are very noticeable, while for the GAR$(2)$, the effect is not as noticeable. In both cases, the improvement obtained by using a burn-in of size 5{,}000 compared to 3{,}000 is small under all metrics.
\begin{figure}
    \centering
    \includegraphics[width=\textwidth]{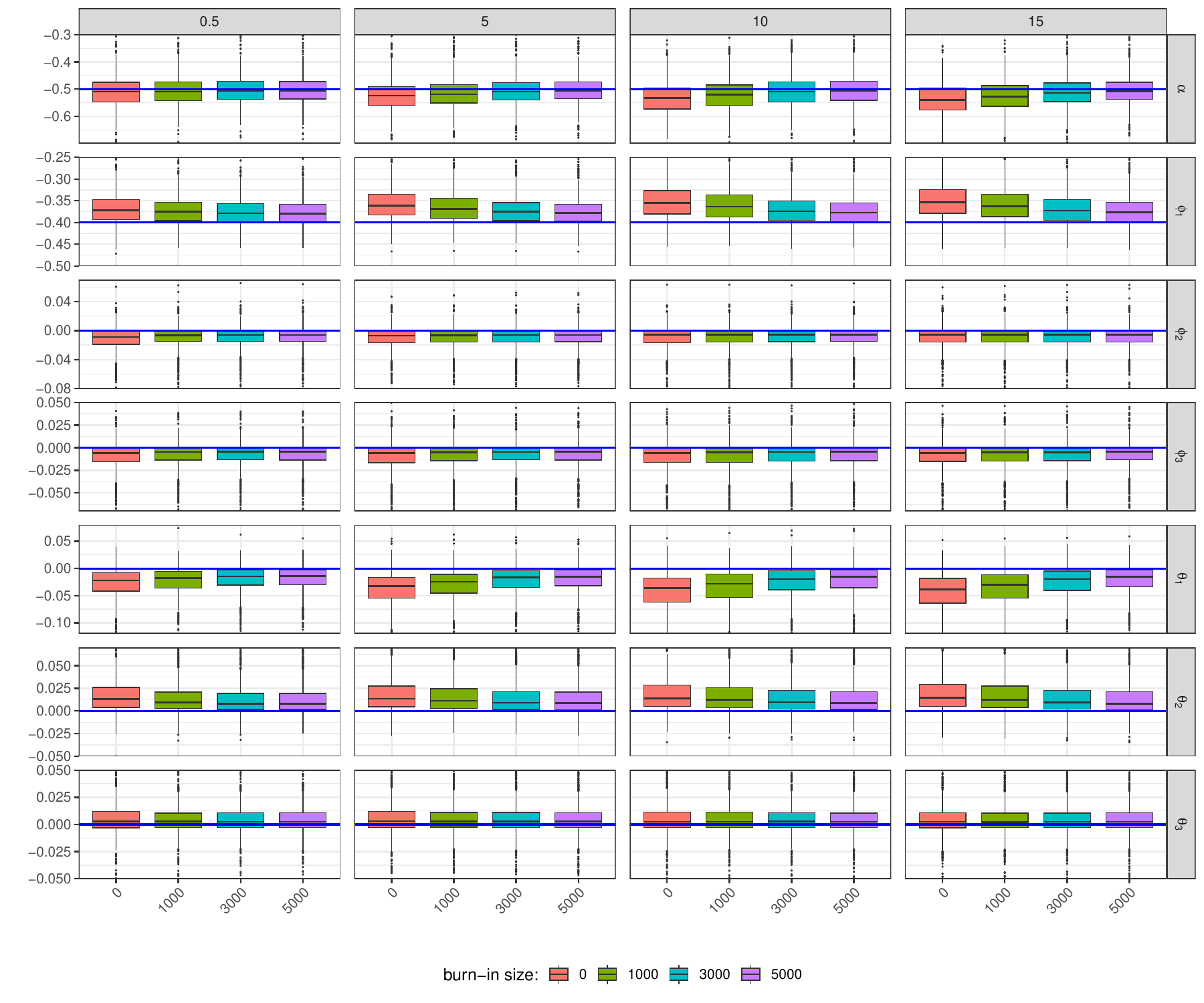}
    \caption{Simulation Results for GAR$(1)$ Model. Presented are the boxplots of point estimates (posterior distribution's average) obtained for each parameter (rows), $\sigma$ (columns) and burn-in sizes (cells).}
    \label{fig:table1}
\end{figure}
\FloatBarrier
\begin{figure}
    \centering
    \includegraphics[width=\textwidth]{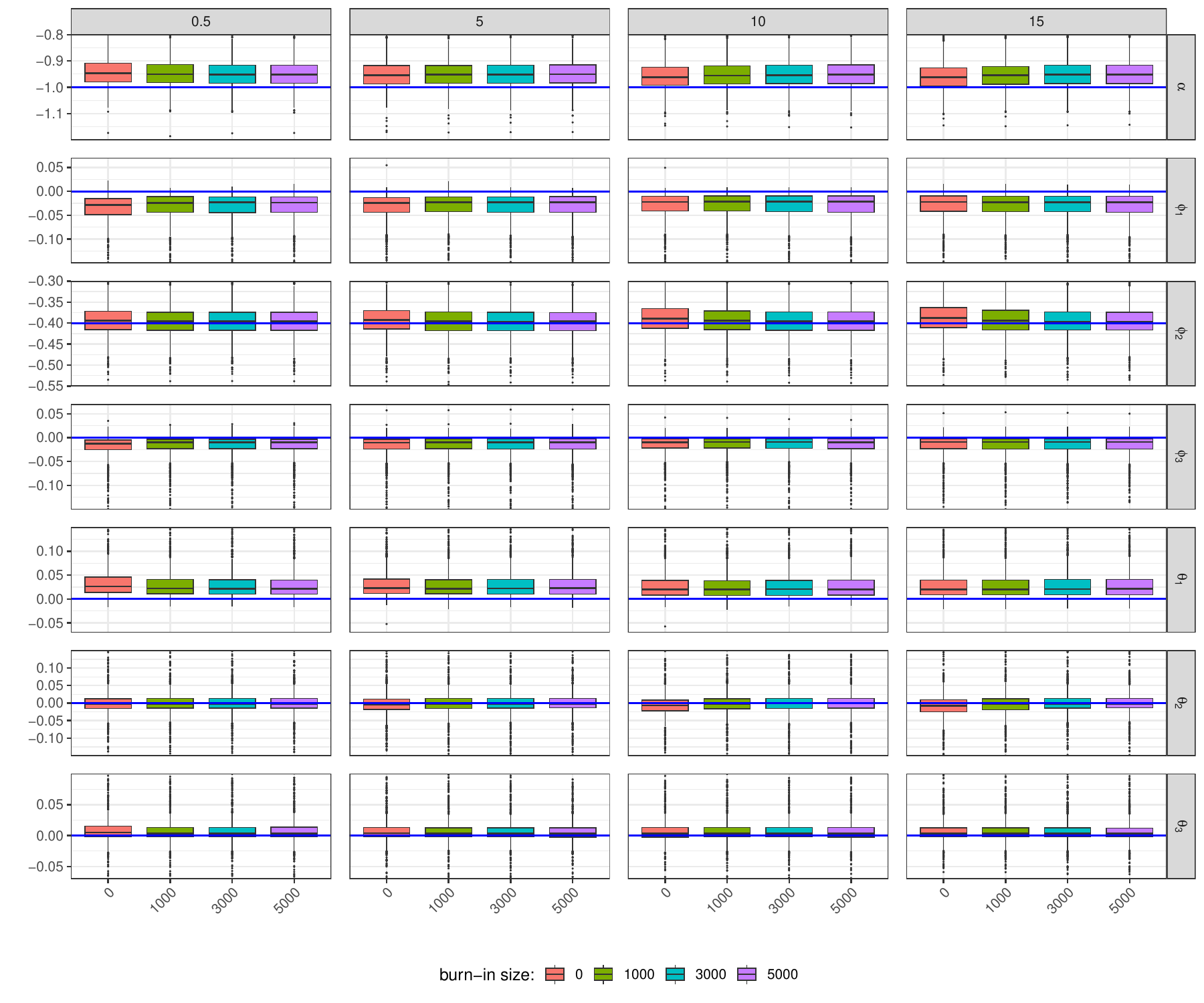}
    \caption{Simulation Results for GAR$(2)$ Model. Presented are the boxplots of point estimates (posterior distribution's average) obtained for each parameter (rows), $\sigma$ (columns) and burn-in sizes (cells). }
    \label{fig:table2}
\end{figure}
\FloatBarrier
\FloatBarrier

%%%%%%%%%%%%%%%%%%%%%%%%%%%%%%%%%%%%%%%%%%%%%%%%%%%%%%%%%%%%%%%%%%%%%%%%%%%%%%%%%%%%%%%%%%%%%%%%%%%%%%%%%%%%%%%%%%%%%%
%%%%%%%%%%%%%%%%%%%%%%%%%%%%%%%%%%%%%%%%%%%%%%%%%%%%%%%%%%%%%%%%%%%%%%%%%%%%%%%%%%%%%%%%%%%%%%%%%%%%%%%%%%%%%%%%%%%%%%
\subsubsection{GMA$(q)$ models}\label{gmaq}
%%%%%%%%%%%%%%%%%%%%%%%%%%%%%%%%%%%%%%%%%%%%%%%%%%%%%%%%%%%%%%%%%%%%%%%%%%%%%%%%%%%%%%%%%%%%%%%%%%%%%%%%%%%%%%%%%%%%%%
In this section we consider GMA$(1)$ models with parameters $\alpha=-0.5$ and $\theta_1=-0.5$ and GMA$(2)$ with $(\alpha, \theta_1, \theta_2)=(-1, 0, 0.6)$. Hyperparameter $m$ was set to 40 and we consider $c=0.3$ for the binomial. To generate the required time series, a burn-in of 100 points was applied yielding a final sample size of  $n=1{,}000$. We generate 1,000 replicas of each proposed scenario.

Regarding the RJMCMC procedures, they are the same as in the previous analysis, namely, maximum orders were taken as $p_m = 3$ and $q_m = 3$, accompanied by a non-informative prior probability of 0.5 for the inclusion of each parameter. Priors for $\alpha$ and the MA parameters were $N(0,0.3^2)$ and $N(0, 0.2^2)$ respectively, whereas $\sigma\in\{0.5, 5,10,15\}$. In each replica, a single chain of 30,000 iterations was sampled for each scenario.

The results are presented in Figures \ref{fig:table3} and \ref{fig:table4} below, Table \ref{burn-inGMA1} and \ref{burn-inGMA2} in the Appendix, and Figure \ref{fig:cov_bi}. Regarding the hyperparameter $\sigma$, in both scenarios $\sigma=0.5$ yielded the worst results, whereas little difference in point estimation is observed for $\sigma\in\{5, 10, 15\}$. Overall, the effects of the burn-in in point estimation are considerably less noticeable than in the GAR case. Considering model identification, in most cases applying a burn-in is even slightly detrimental, especially in the GMA$(1)$ case, {as clearly  seen in the third and fourth rows of Figure \ref{fig:cov_bi}}. The percentage of correctly identified models is lower for the GMA$(1)$ model compared to the GMA$(2)$ model. For GMA$(1)$, model identification performance is slightly higher when using HPD, whereas for GMA$(2)$, no clear pattern is present. Overall, in the GMA situation, applying a burn-in does not seem to significantly improve the results.
\FloatBarrier
\begin{figure}
    \centering
    \includegraphics[width=\textwidth]{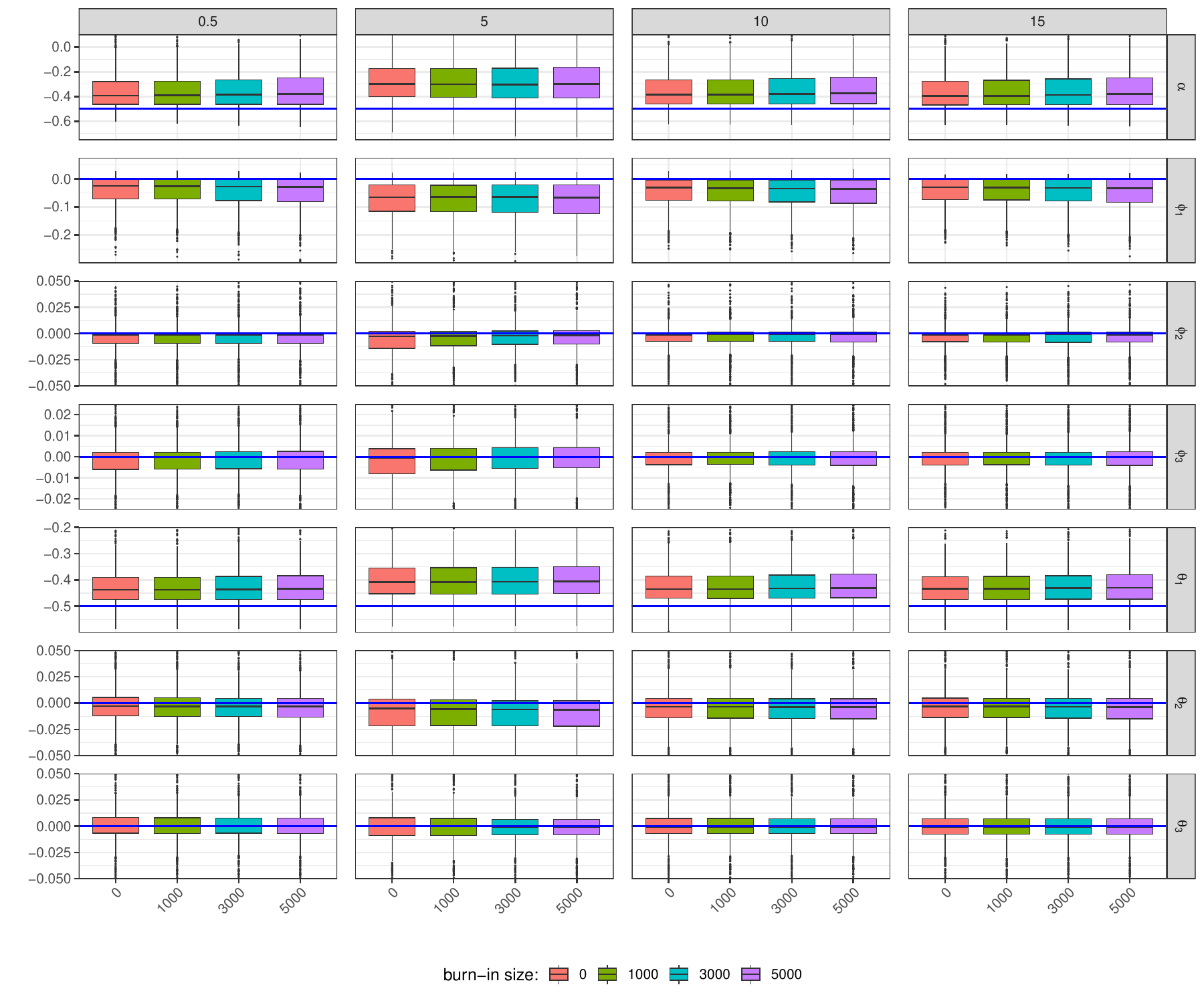}
    \caption{Simulation Results for GMA$(1)$ Model. Presented are the boxplots of point estimates (posterior distribution's average) obtained for each parameter (rows), $\sigma$ (columns) and burn-in sizes (cells). }
    \label{fig:table3}
\end{figure}
\begin{figure}
    \centering
    \includegraphics[width=\textwidth]{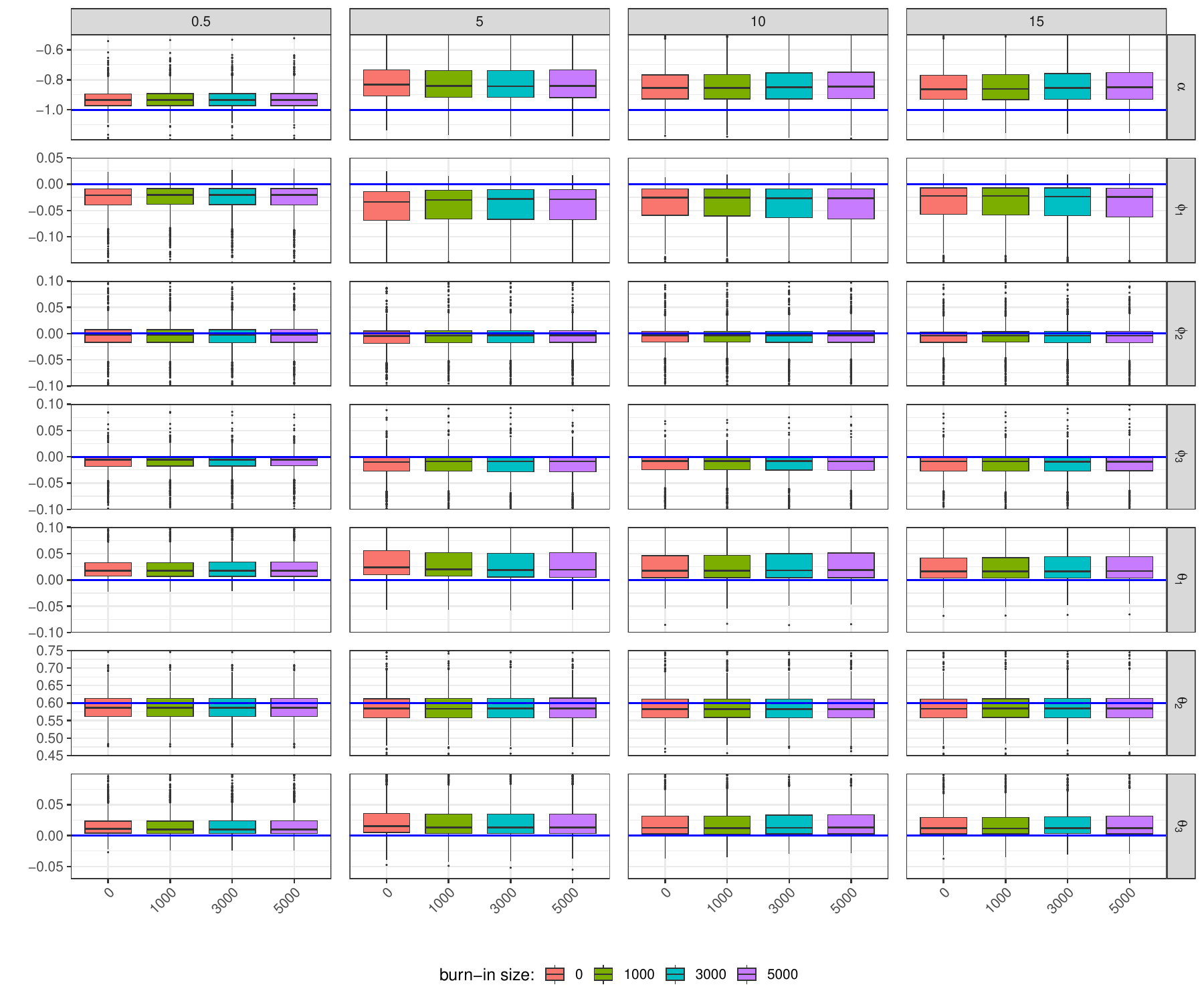}
    \caption{Simulation Results for GMA$(2)$ Model. Presented are the boxplots of point estimates (posterior distribution's average) obtained for each parameter (rows), $\sigma$ (columns) and burn-in sizes (cells). }
    \label{fig:table4}
\end{figure}
\FloatBarrier
\subsection{Effects of Thinning}\label{ET}
Thinning is a technique usually applied when a sample presents considerable autocorrelation. In this section we evaluate the effects of thinning in terms of point and interval estimation, as well as in the effective sample size of each parameter and model identification. The model parameters and other details are kept the same as in Section \ref{BI}. Also notice that no burn-in was applied in this exercise, so that results when no thinning is applied correspond to the case of no burn-in in the previous section.
\subsubsection{GAR$(p)$ models}
\begin{figure}
    \centering
    \includegraphics[width=\textwidth]{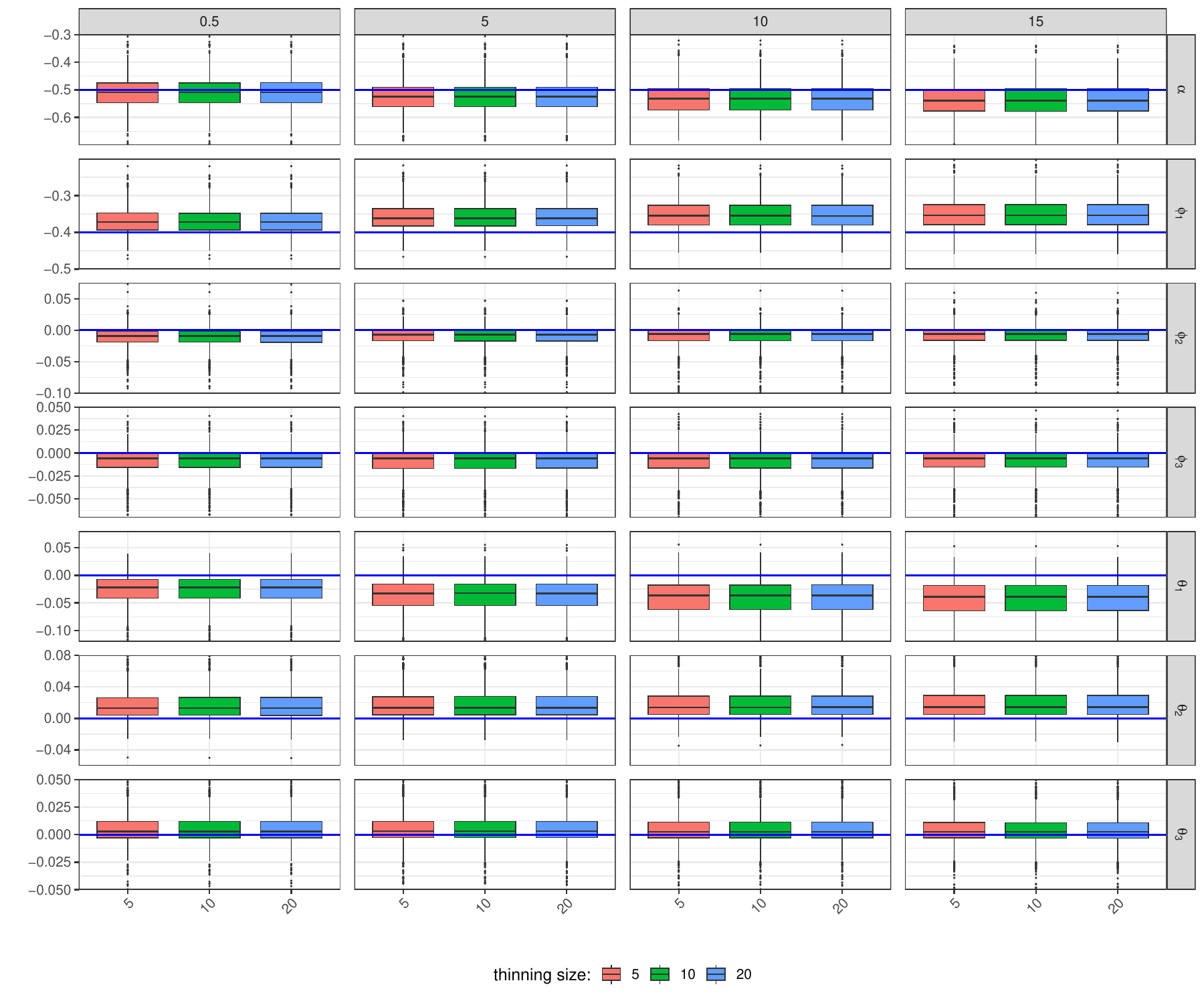}
    \caption{Simulation Results for GAR$(1)$ Model. Presented are the boxplots of point estimates (posterior distribution's average) obtained for each parameter (rows), $\sigma$ (columns) and thinning sizes (cells). }
    \label{fig:table_thinning1}
\end{figure}

Considering the GAR$(1)$ and GAR$(2)$ models presented in Section \ref{BI}, we now study the effects of applying  thinning of lags $\{5,10,20\}$ in the posterior samples prior to inference. The case of no thinning corresponds to the case of no burn-in presented in the previous section. The results are presented in {Figures \ref{fig:table_thinning1} and \ref{fig:table_thinning2} below} and Tables \ref{thinningGAR1} and \ref{thinningGAR2} presented in the Appendix.  Regarding point estimation, applying any thinning does not improve the results in any way. This is expected since both the sample mean and sample median are consistent estimator even under dependence in the data. Hence, even when applying a thinning of 20, the final sample is of size 1,500, which is still sufficiently large to guarantee that the sample mean and sample median are very close to the ones obtained with no thinning. Similar reasoning apply to the construction of credibility intervals, which in turn imply that thinning is expected to have little impact on model selection. These results are all reasonable considering that thinning is mainly used to reduce the correlation in the sample improving effective sample size. So, does effective sample size values improve after application of the thinning? Well, not quite. The simulation results shown borderline improvements at best, and even some decline in a few cases, especially for the GAR$(2)$ model (see Tables \ref{thinningGAR1} and \ref{thinningGAR2} in the Appendix).
%
% \begin{table}
% \centering
% \renewcommand{\arraystretch}{1.2}
% \setlength{\tabcolsep}{6pt}
% \caption{Correctly identification percentage for HPD and ECI for GAR$(1)$ and GAR$(2)$ models with thinning $\{0, 10, 20\}$ and $\sigma\in\{0.5,5,10,15\}$.}\vspace{.3cm}\label{cov_GAR_th}
% \begin{tabular}{c|c|cc|cc|cc|cc}
% \hline\hline
% \multirow{2}{*}{Model}& \multirow{2}{*}{Thinning}&  \multicolumn{2}{c}{$\sigma = 0.5$} &  \multicolumn{2}{c}{$\sigma = 5$}     &  \multicolumn{2}{c}{$\sigma = 10$} &  \multicolumn{2}{c}{$\sigma = 15$} \\
%  \cline{3-10}
%     &                     &               HPD    &    ECI        &               HPD    &    ECI        &               HPD    &    ECI      &               HPD    &    ECI      \\
% \hline\hline\multirow{3}{*}{GAR$(1)$}
% &5    & 74.0\%  &  66.5\%  &  55.7\%  & 48.2\%  & 47.7\%  &  40.9\%  & 44.4\%   & 37.5\% \\
% \cline{2-10}
% &10 & 73.7\%    &  66.6\%  &  55.7\%  &  48.2\% & 47.7\%  & 40.9\%   &  44.2\%  & 37.5\% \\
% \cline{2-10}
% &20   & 73.8\%  & 66.8\%   & 55.9\%   &  48.4\% & 47.6\%  & 41.3\%   & 44.3\%   & 37.7\% \\
% \hline\hline \multirow{3}{*}{GAR$(2)$}
% &5    & 98.7\%  & 97.2\%  & 95.9\%  & 89.2\%  & 92.1\%  & 81.5\%  & 87.2\%  & 76.8\% \\
% \cline{2-10}
% &10   & 98.7\%  & 97.2\%  & 95.8\%  & 89.1\%  & 92.1\%  & 81.6\%  & 87.2\%  & 76.8\% \\
% \cline{2-10}
% &20   & 98.7\%  & 97.4\% &  95.6\%  & 89.5\%  & 92.0\%  & 81.8\%  & 87.0\%  & 77.2\% \\
% \hline\hline
% \end{tabular}
% \end{table}
% %
\begin{figure}
    \centering
    \includegraphics[width=\textwidth]{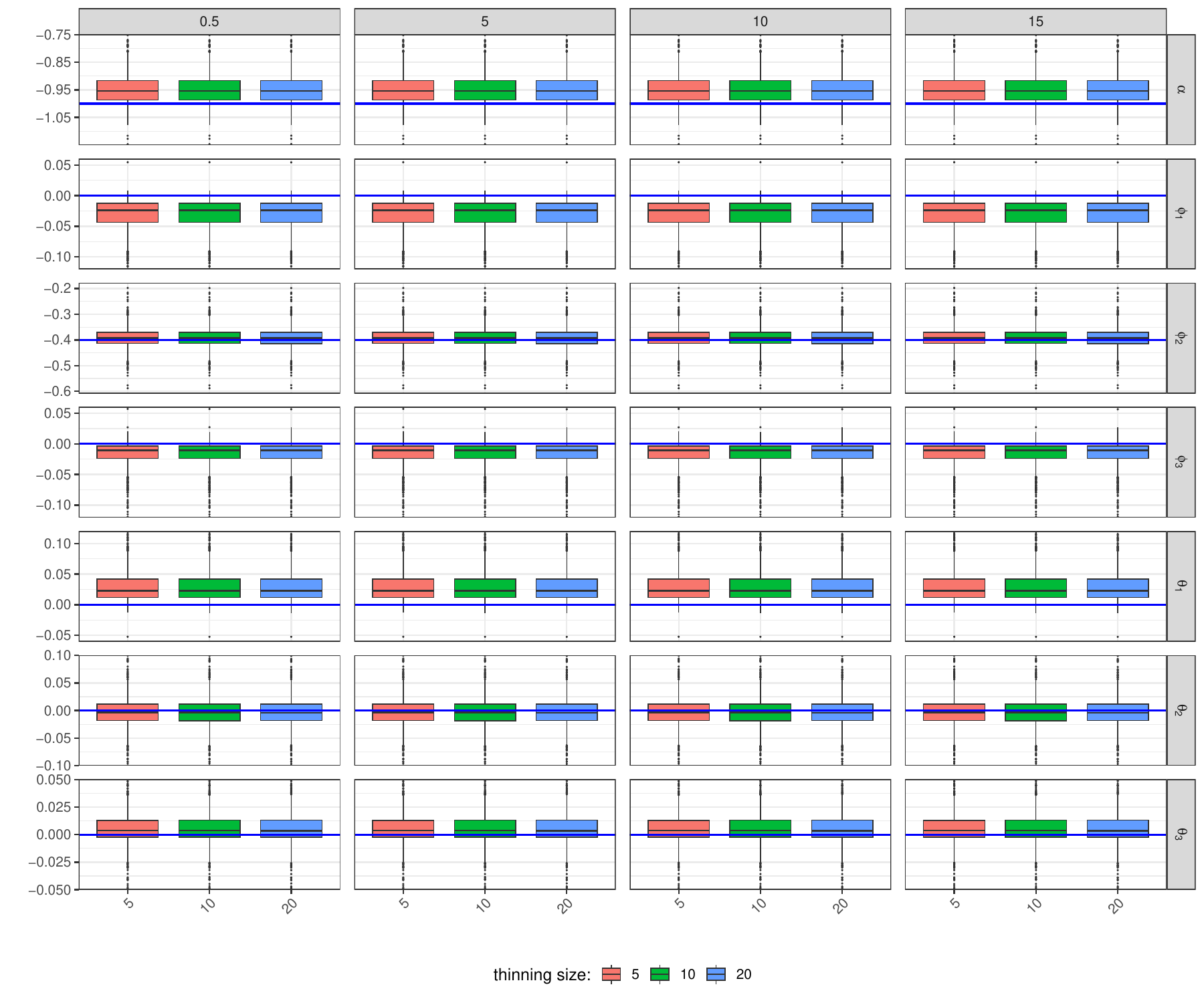}
    \caption{Simulation Results for GAR$(2)$ Model. Presented are the boxplots of point estimates (posterior distribution's average) obtained for each parameter (rows), $\sigma$ (columns) and thinning sizes (cells).  }
    \label{fig:table_thinning2}
\end{figure}
\FloatBarrier
\subsubsection{GMA$(q)$ models}
\begin{figure}
    \centering
    \includegraphics[width=\textwidth]{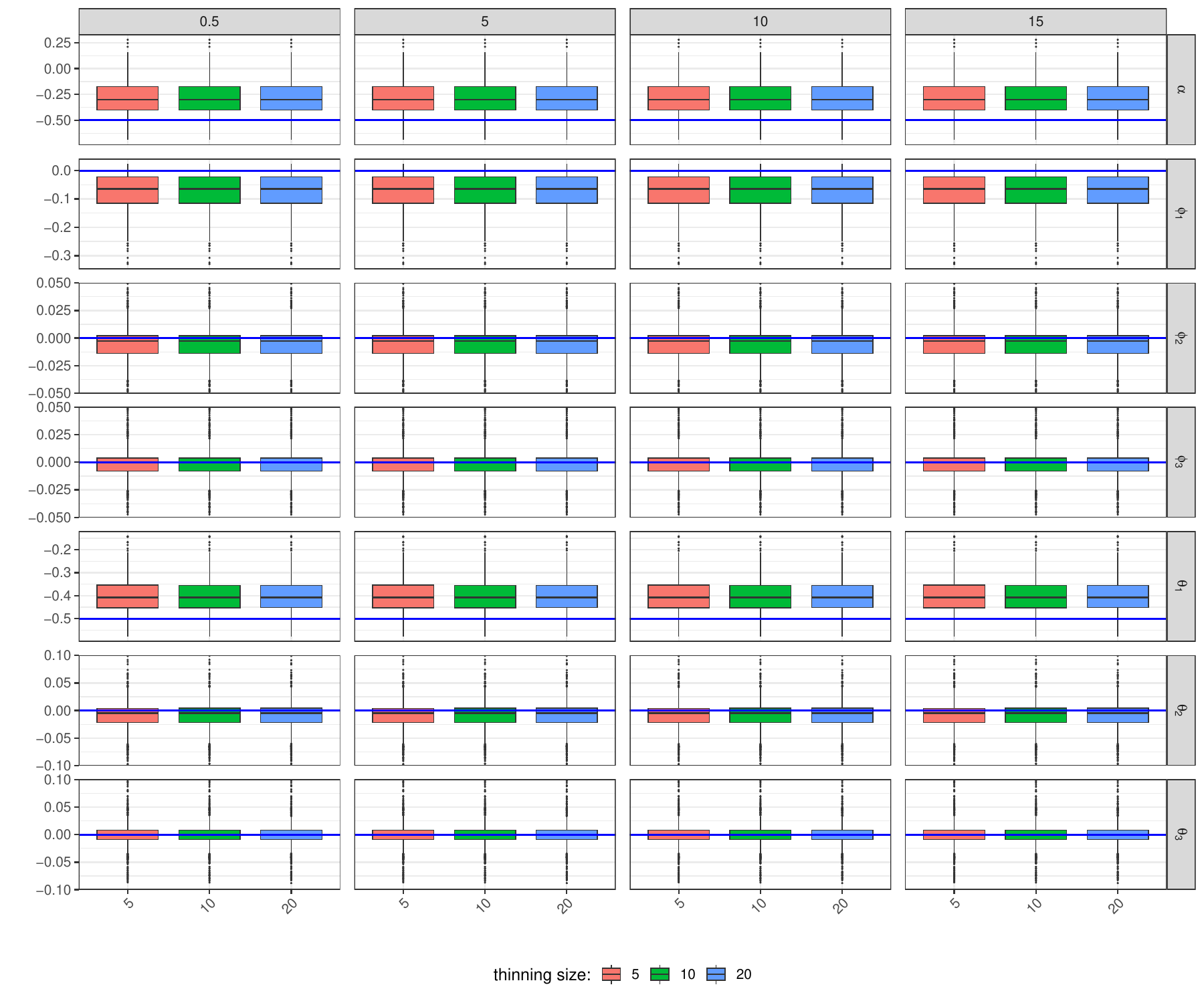}
    \caption{Simulation Results for GMA$(1)$ Model. Presented are the boxplots of point estimates (posterior distribution's average) obtained for each parameter (rows), $\sigma$ (columns) and thinning sizes (cells).}
    \label{fig:table_thinning3}
\end{figure}

{Considering the GMA$(1)$ and GMA$(2)$ models presented in Section \ref{BI}, Figures \ref{fig:table_thinning3} and \ref{fig:table_thinning4} below and Tables \ref{thinningGMA1} and \ref{thinningGMA2} in the Appendix,} display the simulation results obtained by applying a thinning of size $\{5,10,20\}$ prior to inference. Analogously to the results for GAR models, applying a thinning did not present a significant effect on point estimates, and little to no improvement in the effective sample size.
\subsubsection{Summary}
In summary, considering the scenarios studied in the paper, we have evidence that the use of a burn-in before proceeding with inference is very effective in improving point estimation and model selection, whereas using a thinning approach does not significantly improve effective sample size. Furthermore, the use of a burn-in mitigates the dependence in the scale hyperparameter otherwise observed in the results, allowing for a more reliable use of the method.
%
% \begin{table}
% \centering
% \renewcommand{\arraystretch}{1.2}
% \setlength{\tabcolsep}{6pt}
% \caption{Correctly identification percentage for HPD and ECI for GMA$(1)$ and GMA$(2)$ models with thinning $\{0, 10, 20\}$ and $\sigma\in\{0.5,5,10,15\}$.}\vspace{.3cm}\label{cov_GMA_th}
% \begin{tabular}{c|c|cc|cc|cc|cc}
% \hline\hline
% \multirow{2}{*}{Model}& \multirow{2}{*}{Thinning}&  \multicolumn{2}{c}{$\sigma = 0.5$} &  \multicolumn{2}{c}{$\sigma = 5$}     &  \multicolumn{2}{c}{$\sigma = 10$} &  \multicolumn{2}{c}{$\sigma = 15$} \\
%  \cline{3-10}
%     &                     &               HPD    &    ECI        &               HPD    &    ECI        &               HPD    &    ECI      &               HPD    &    ECI      \\
% \hline\hline\multirow{3}{*}{GMA$(1)$}
% &5    & 89.3\%  & 88.4\%  & 94.0\%  & 93.7\%  & 94.9\%  & 95.0\%  & 95.7\%  & 95.3\% \\
% \cline{2-10}
% &10   & 89.2\%  & 88.5\%  & 94.1\%  & 93.7\%  & 95.0\%  & 95.0\%  & 95.7\%  & 95.2\% \\
% \cline{2-10}
% &20   & 89.3\%  & 88.3\%  & 94.0\%  & 93.7\%  & 95.1\%  & 95.0\%  & 95.8\%  & 95.2\% \\
% \hline\hline \multirow{3}{*}{GMA$(2)$}
% &5    & 98.8\%  & 99.1\%  & 99.4\%  & 99.4\%  & 99.2\%  & 99.2\%  & 99.0\%  & 98.8\% \\
% \cline{2-10}
% &10   & 98.7\%  & 99.1\%  & 99.4\%  & 99.4\%  & 99.2\%  & 99.2\%  & 99.0\%  & 98.8\% \\
% \cline{2-10}
% &20   & 98.7\%  & 99.1\%  & 99.5\%  & 99.4\%  & 99.2\%  & 99.2\%  & 99.0\%  & 98.8\% \\
% \hline\hline
% \end{tabular}
% \end{table}
%
%
\FloatBarrier
\begin{figure}
    \centering
    \includegraphics[width=\textwidth]{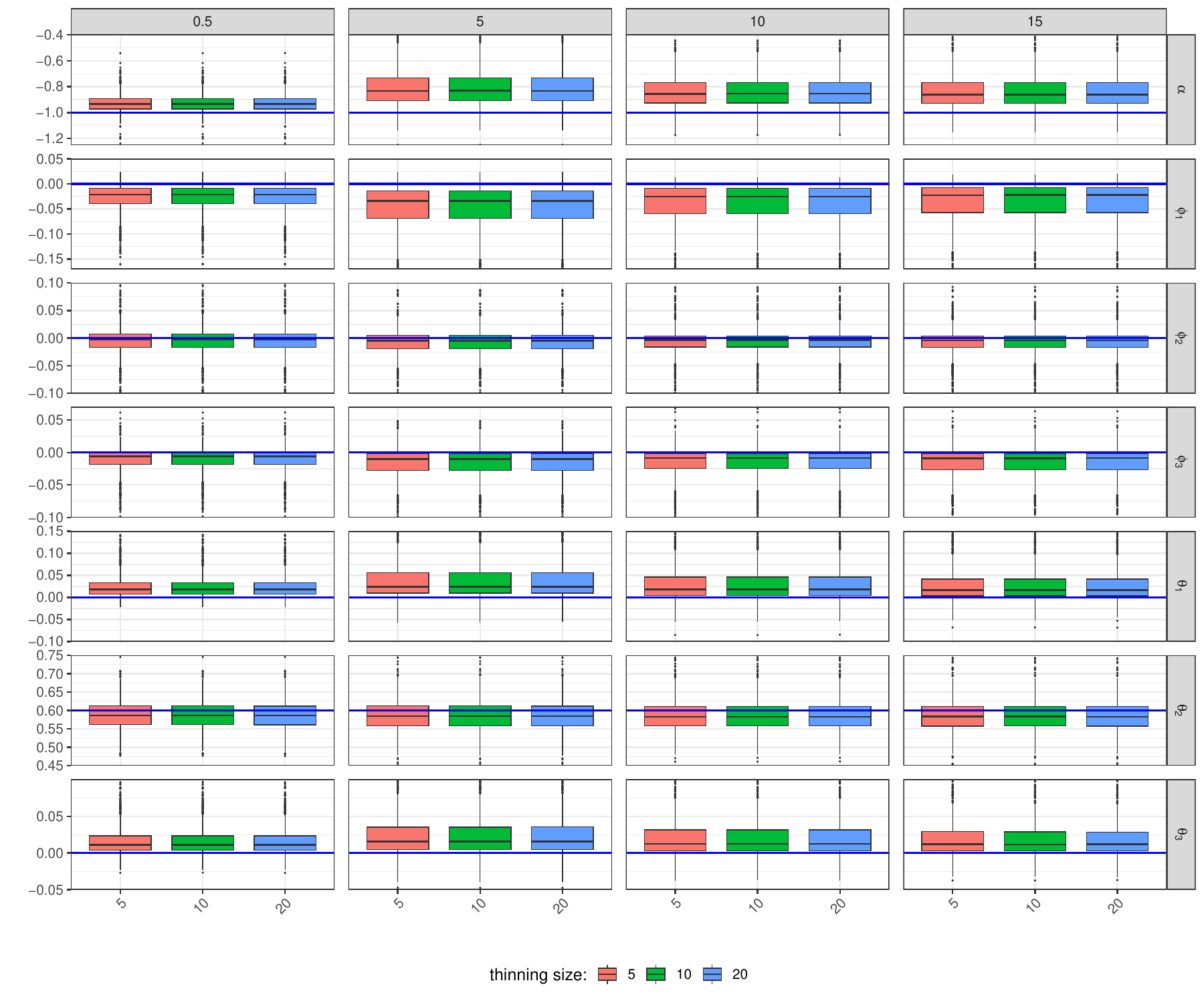}
    \caption{Simulation Results for GMA$(1)$ Model. Presented are the boxplots of point estimates (posterior distribution's average) obtained for each parameter (rows), $\sigma$ (columns) and thinning sizes (cells).  }
    \label{fig:table_thinning4}
\end{figure}
\FloatBarrier
{
\subsection{Average runtime}
In this section, we provide information on the average runtime required to generate a single chain of size 30{,}000 using the proposed methodology. Specifically, we address the following questions: Does the time required to generate the chains vary with the model type? How does the scale parameter $\sigma$ affect the runtime?

To answer these questions, we conducted a small simulation considering the same setups presented in Sections \ref{garp} and \ref{gmaq}. For each scenario, 10 replicates were generated, and the \texttt{R} function \texttt{system.time} was used to measure the elapsed time for generating each chain. The simulations were run sequentially on a PC using \texttt{R} version 4.0.3, with the following specifications: Intel Core i5-8600k CPU (3.6 GHz, factory settings), 16 GB RAM, and Windows 10 Pro.

The results of the simulation study are summarized in Table \ref{runtime}, which displays  the mean and standard deviation (sd) of the runtime (in seconds) for generating a single chain containing 30,000 iterations for four model configurations: GAR$(1)$, GAR$(2)$, GMA$(1)$, and GMA$(2)$, across four values of the scale parameter $\sigma$: 0.5, 5, 10, and 15. Overall, the runtime remains relatively consistent across the different scale parameter values, with slight variations. In all cases, scale $\sigma=0.5$ took between 2.9\% (GAR$(2)$) to 9.5\% (GMA$(1)$) longer to run on average, than $\sigma=15$. For GMA$(1)$, the mean runtime decrease slightly as $\sigma$ increases. The standard deviations indicate moderate variability in runtime, with GMA$(2)$ exhibiting slightly larger variations compared to the other configurations. These findings suggest that while the scale parameter $\sigma$ and model type may influence runtime slightly, the overall differences are minimal.
\begin{table}[ht!]
\centering
\renewcommand{\arraystretch}{1.2}
\setlength{\tabcolsep}{6pt}
\caption{Mean runtime (seconds) and standard deviation (sd) for generating a chain of size 30,000 across scale parameter values ($\sigma$) and model configurations.}\vspace{.3cm}\label{runtime}
\begin{tabular}{c|cc|cc|cc|cc}
\hline
\multirow{2}{*}{Scale} & \multicolumn{2}{c|}{GAR$(1)$}& \multicolumn{2}{c|}{GAR$(2)$} & \multicolumn{2}{c|}{GMA$(1)$} & \multicolumn{2}{c}{GMA$(2)$}\\
\cline{2-9}
   & mean & sd & mean & sd & mean & sd & mean & sd \\
         \hline
0.5  &142.6 &5.25  &137.4 &6.78  &148.0 & 5.75  &  142.9 & 8.42   \\
5    &131.7 &6.91  &138.2 &7.29  &146.0 & 5.64  &  148.8 & 6.02    \\
10   &138.0 &5.54  &132.7 &5.79  &136.8 & 4.33  &  136.0 & 8.03   \\
15   &137.5 &4.29  &133.5 &5.42  &135.2 & 5.94  &  137.8 & 7.74   \\
\hline
\end{tabular}
\end{table}
}
\FloatBarrier
\section{Applications}
In this section we present two illustrative applications of the proposed methodology highlighting its potential in model selection under different scenarios. The first application involves automobile production in Brazil and demonstrates how the methodology can be used for model selection in the context of count time series, including long-term trend selection.  The second application is related to bus production in Brazil before and after the COVID pandemics, illustrating how to apply the methodology to conduct a pre/post-event analysis of count time series.

\subsection{Automobile production in Brazil}

In this section we present an application of the proposed methodology to analyze the automobile production in Brazil between January 1993 and December 2013, which yielding a sample size of $n=252$ observations. The same data was considered in \cite{bgarma}. As in the mentioned work, the data was divided by 1,000 to reduce its magnitude and rounded to the nearest integer when necessary. The data is freely available from the ANFAVEA (the Brazilian National Association of Motor Vehicle Manufacturers) website: \url{http://www.anfavea.com.br}. In \cite{bgarma}, the authors fit a negative binomial GARMA$(1,1)$ model to the data under a Bayesian framework. We are particularly interested in model selection, conducted using information criteria as guideline in the aforementioned paper. Instead, we shall conduct model selection using the proposed RJMCMC approach.

\begin{figure}
    \centering
    \includegraphics[width=0.45\textwidth]{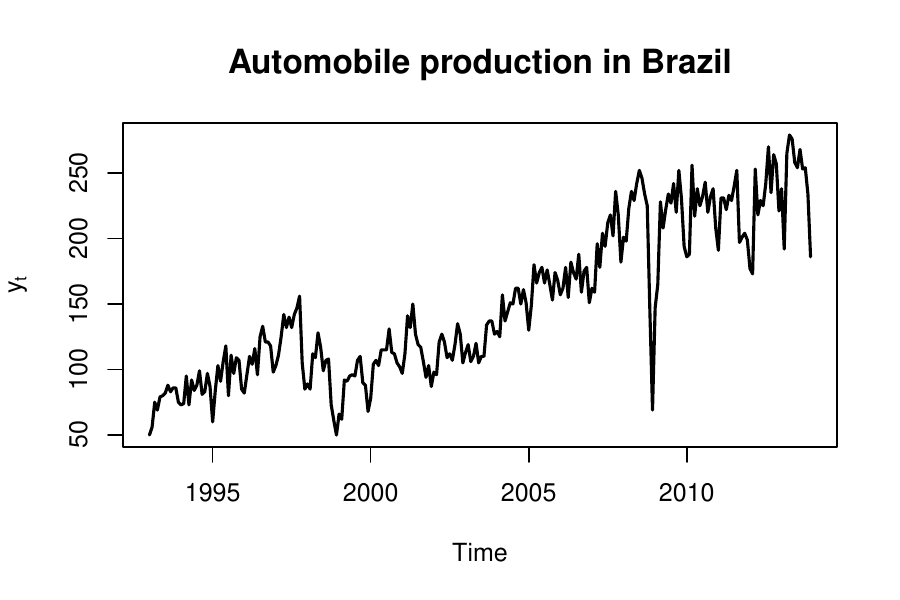}
    \includegraphics[width=0.45\textwidth]{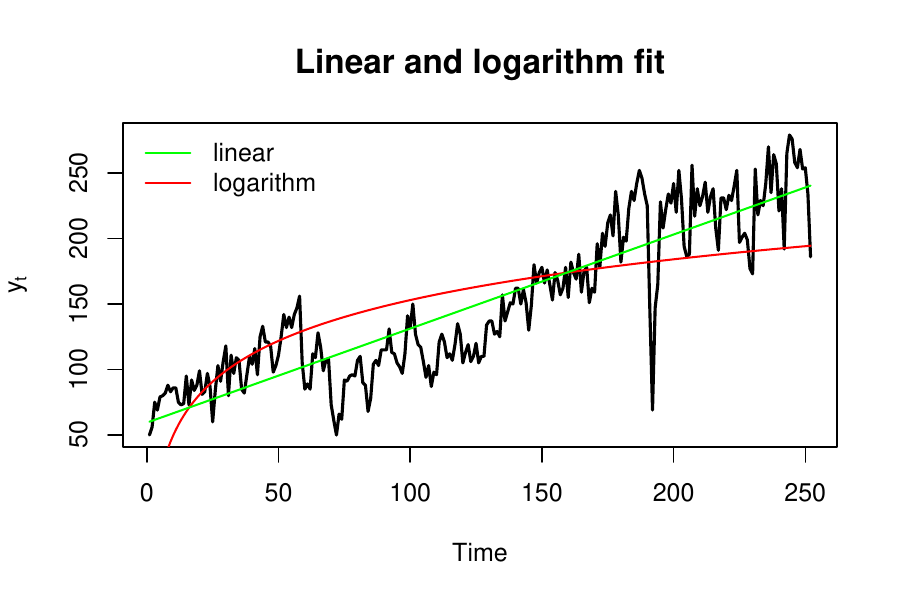}
    \caption{Automobile production in Brazil from January 1993 to December 2013. Shown are the time series plot alone in the left and along the fitted linear and logarithm trends  on the right. }
    \label{fig:etiqueta}
\end{figure}

The time series plot is presented in Figure \ref{fig:etiqueta} (left) and reveals the presence of a visible increasing trend. To account for this, \cite{bgarma} considered a logarithmic trend as covariate in the model. However, considering the data directly, {simple visual inspection clearly indicates that a linear trend provides a better fit. This can also be confirmed by a simple regression model.} Let $y_1,\cdots, y_n$ denote the observed time series. We fit the following linear models to the data:
\begin{equation*}
\mathrm M1:\ y_t = a_0+a_1\log(t)+e_t\quad\mbox{and}\quad \mathrm M2: \ y_t = b_0+b_1t+e_t,
\end{equation*}
where $e_t$ denotes a generic error term. The ordinary least squares estimates of the models are $\hat a_0=-53.74$, $\hat a_1=44.89$, $\hat b_0=59.41$ and $\hat b_1=0.72$. The time series plot along with the fitted values for M1 and M2 are shown in Figure \ref{fig:etiqueta} (right). For M1, $R^2=0.54$, with residual standard error of 40.06, while for M2, $R^2=0.79$ with a residual standard error of 26.9. These results favor the linear trend as a better fit for the long-term growth observed in the time series. However, since GARMA models are defined in a GLM fashion, the linear trend may not outperform the logarithmic trend when the GARMA structure is considered. To determine which trend is more appropriate to model the data, we will embed the trend term into the RJMCMC strategy, incorporating trend selection along with model selection.
The most complex GARMA$(p_m,q_m)$ we will consider consists of random component given by \eqref{BNModel1} along with systematic component given by
\begin{equation}\label{app}
\log(\mu_t)=\beta_0+\beta_1t+\beta_2\log(t)+\sum_{j=1}^{p_m} \phi_{j}\big[\log (Y_{t-j}^{*})-\beta_1(t-j)-\beta_2\log(t-j)\big]+\sum_{j=1}^{q_m} \theta_{j}r_{t-j},
\end{equation}
with $r_t:=\log(Y^{*}_{t})-\log(\mu_{t})$ and $Y^{*}_t=\max\{0.3,Y_t\}$, that is, we set $c=0.3$. Parameter $\beta_0$ is always included in the sampler, while all other parameters are targets for model transition. We found that convergence is very slow in this scenario, so the RJMCMC is configured to produce a single chain containing 300,000 iterations, with the first 295,000 are discarded as burn-in. The scale hyperparameter is set to 5, $m=150$ just as in \cite{bgarma} and the inclusion probability for each parameter is set to 0.5. All parameters are initialized in \texttt{Nimble} as 0. The prior distributions are given by: $\beta_0\sim N(0,0.3^2)$, $\phi_i\sim N(0,0.2^2)$, $\theta_i\sim N(0,0.2^2)$, $\beta_j\sim N(0,16)$, for $i\in\{1,2,3\}$ and $j\in\{1,2\}$.

\begin{figure}[!hb]
    \centering
    \includegraphics[width=0.8\textwidth]{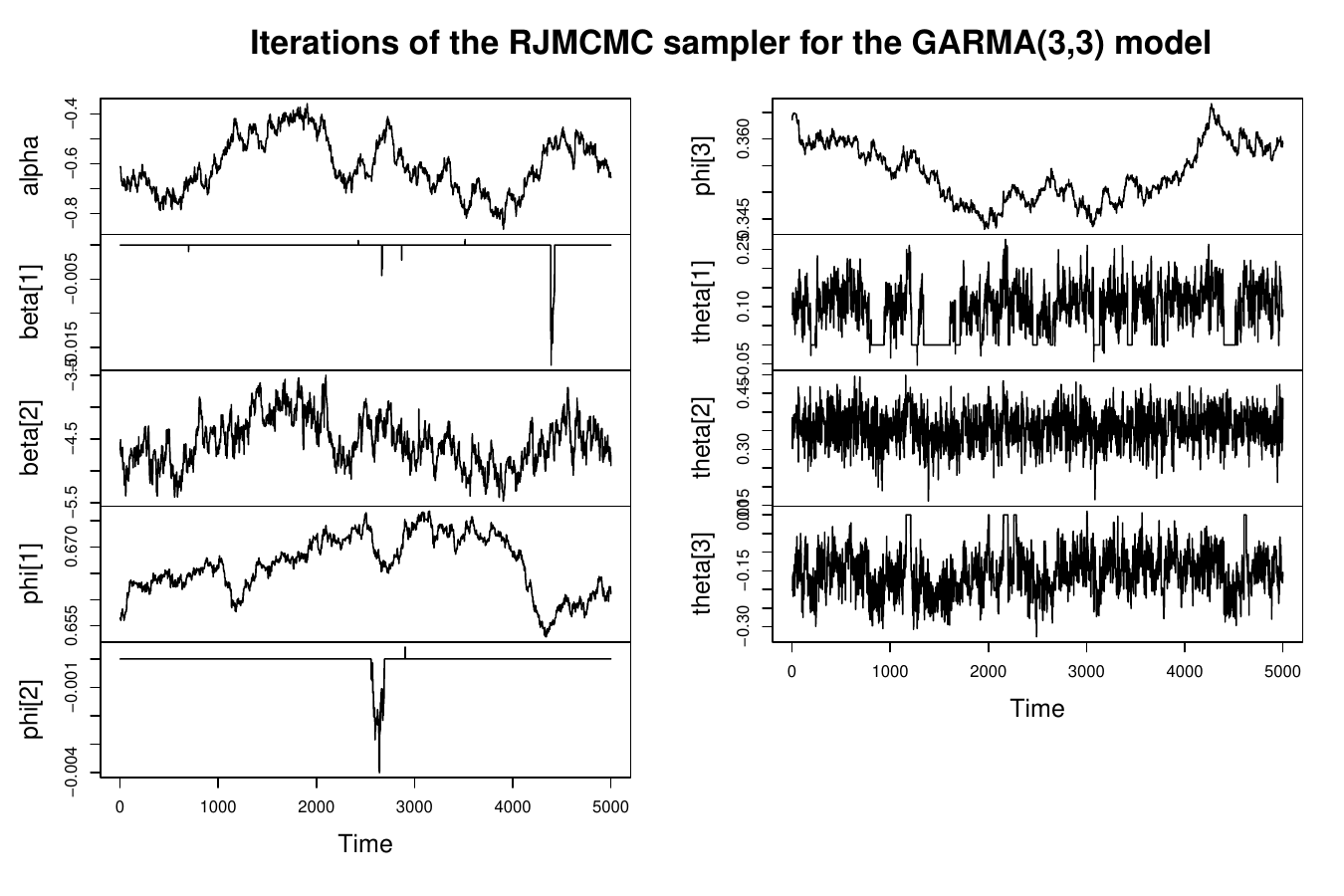}
    \caption{Iterations of the RJMCMC sampler. }
    \label{iter}
\end{figure}

The first exercise involves setting $p_m=q_m=3$ and running the RJMCMC. The results are presented in Table \ref{M33} and the time series plot of the generated chain is shown in Figure \ref{iter}. The last column of Table \ref{M33} presents Geweke's convergence diagnostic (GCD), which tests the equality of the means of the first 10\% and last 50\% of a Markov chain \citep{gcd}. The displayed values are the $z$-scores calculated under the assumption that the two parts of the chain are asymptotically independent. We observe that all values are smaller than 1.96 in absolute value, indicating that the chain of each parameter converged to its target distribution at a 95\% confidence level.
From the point estimates, the first thing we notice is that the RJMCMC selected the logarithm long-term growth, excluding $\beta_1$ from the model in almost all iterations. This also occurs with parameters $\phi_2$, which is nearly all iterations. Besides $\beta_1$ and $\phi_2$, $\theta_1$ is also non-significant at a 95\% confidence level HPD credibility interval, although it was frequently selected for inclusion in the model. All other parameters can be considered significant according to the HPD credible interval. Median and mean estimates are very close indicating symmetry of the target distribution. The roots of the characteristic polynomial for the AR component are all greater than 1.236, thus lying outside of the unit circle.

\begin{figure}[!ht]
    \centering
    \includegraphics[width=0.6\textwidth]{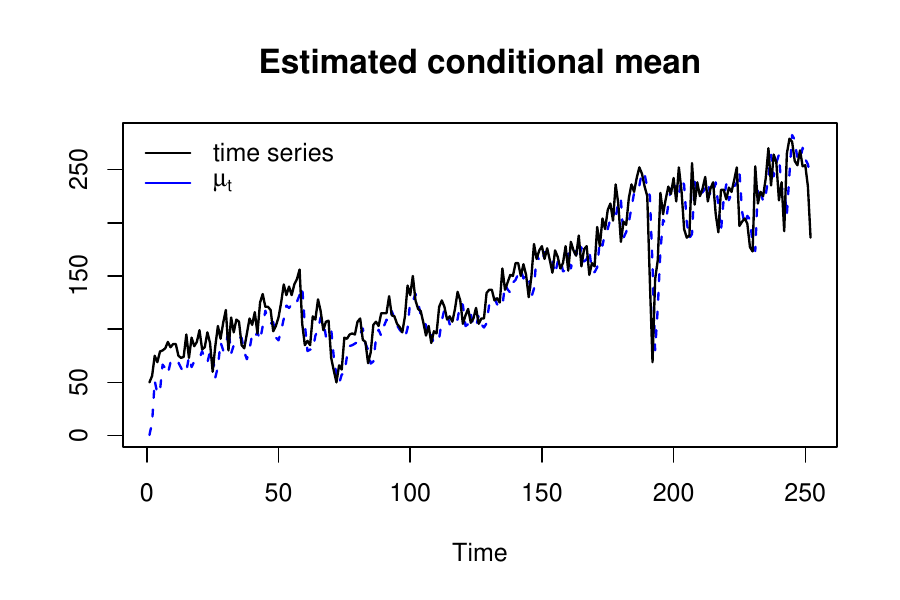}
    \caption{Reconstructed conditional mean. }
    \label{mutt}
\end{figure}

In Figure \ref{mutt}, we present the reconstructed conditional mean $\mu_t$ based on the (mean) estimated values along with the original time series. This seemingly delayed pattern is commonly seen in GARMA models containing autoregressive components. As expected, $\mu_t$ accompanies $y_t$ very closely, indicating that the model is a good fit.
In \cite{bgarma} based on information criteria, the authors selected a NB-GARMA$(1,1)$ model with a logarithm trend as the best model among those considered. Using the proposed RJMCMC approach, we selected a more complex model,  technically a NB-GARMA$(3,3)$, but with coefficients $\phi_2$ and $\theta_1$ equal zero. The method also identified the logarithmic long-term growth as the most appropriate  for the data. Unfortunately, a deeper comparison between our results and those in \cite{bgarma} is not possible due to missing  key information in the mentioned paper. For instance, there is no indication of the value of the constant $c$ applied, nor about the number of iterations and the burn-in period used.

\begin{table}
\small
\centering
\renewcommand{\arraystretch}{1.1}
\setlength{\tabcolsep}{10pt}
\caption{Results from fitting a NB-GARMA$(3,3)$ model defined in \eqref{app}.}\vspace{.3cm}\label{M33}
\begin{tabular}{lcccccc}
\hline
Par  & Mean & Median & SD  & HPD CI (95\%) & GCD \\
\hline
$\beta_0$ & -0.606  & -0.620    & 0.109  &  $  [-0.786, -0.402]  $   & -1.231    \\
$\beta_1$  &0.000    &0.000     & 0.001  &  $  - $                  & 1.229    \\
$\beta_2$  &-4.542   &-4.571    & 0.377  &  $  [-5.203, -3.750] $    & -1.763   \\
$\phi_1$   &0.667    &0.667     & 0.005  &  $  [0.657, 0.676 ] $    & -0.982    \\
$\phi_2$   &0.000    &0.000     & 0.000  &  $ - $                   & 0.696      \\
$\phi_3$   &0.354    &0.353     & 0.005  &  $  [0.344, 0.362] $      & 1.820    \\
$\theta_1$ &0.093    &0.099     & 0.063  &  $  [0.000, 0.196] $     & -0.077   \\
$\theta_2$ &0.358    &0.360     & 0.047  &  $  [0.267, 0.445 ] $     & 0.638    \\
$\theta_3$ &-0.154   &-0.156    & 0.060  &  $  [-0.285, -0.049 ] $   & -0.983   \\
\hline
\end{tabular}
\end{table}
\FloatBarrier

\subsection{Bus exportation in Brazil before and after the COVID-19 pandemic}
\begin{figure}[!ht]
    \centering
    \includegraphics[width=0.6\textwidth]{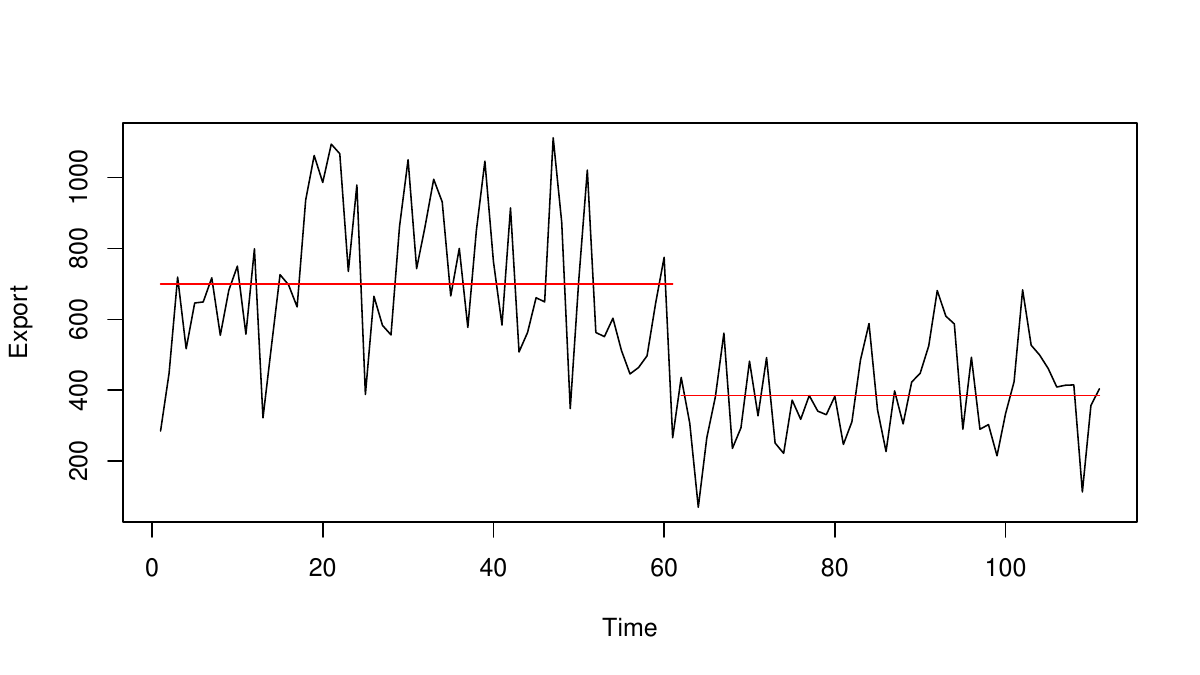}
    \caption{Time series plot of the number of exported buses from Brazil showing the difference in levels before and after February 2020.}
    \label{tsplot}
\end{figure}

In this section we present an application of the proposed methodology to bus exportation in Brazil before and after the COVID-19 pandemic. The data comprises the  monthly number of exported buses as reported by ANFAVEA from January 2015 to March 2024 (as of the first day of each month), yielding a sample size  $n = 111$. Let $y_1,\cdots, y_{111}$ denote the sample. A time series plot reveals a sudden change in level starting in February 2020, as a consequence of the COVID-19 pandemic. The time series plot is shown in Figure \ref{tsplot}. Let $x_t$ be a dummy variable indicating the start of the pandemic's effects in the bus exports, taking value 0 for $t\in\{1,\cdots,61\}$ (up to February 2020) and 1 afterwards. To obtain an idea of the pandemic's effect in the mean exportation of buses from Brazil, a simple regression
\[y_t=\beta_0+\beta_1x_t+\eps_t,\]
fitted using ordinary least squares reveals $\hat\beta_0=699.9$ and $\beta_1=-314,6$ (p-values $<10^{-14}$), indicating that, on average, bus exports decreased by about 314 buses per month due to the pandemic. The fitted values are also presented in Figure \ref{tsplot}. Interestingly, this reduction  persists in a seemingly stationary state after the change in level.

\FloatBarrier
In conclusion, the change in mean exports, which is evident in the plots, is also statistically significant. However, one question remains: is the time series behavior before and after the change in mean the same? To answer this, we propose dividing the time series into two sub-series, one before and other after the change in mean, and applying the proposed reversible jump methodology to each sub-series, comparing the resulting models. Before proceeding with the division, we fit a NB-GARMA model to the full time series, considering $p_m=q_m=3$ and $x_t$ as covariate.  The most complex GARMA$(p_m,q_m)$ considered in the RJ consists of random component given by \eqref{BNModel1} along with systematic component given by
\begin{equation}\label{app2}
\log(\mu_t)=\beta_0+\beta_1x_t+\sum_{j=1}^{p_m} \phi_{j}\big[\log (Y_{t-j}^{*})-\beta_1(t-j)-\beta_2\log(t-j)\big]+\sum_{j=1}^{q_m} \theta_{j}r_{t-j},
\end{equation}
with $r_t:=\log(Y^{*}_{t})-\log(\mu_{t})$ and $Y^{*}_t=\max\{0.3,Y_t\}$. Parameter $\beta_0$ is always included in the sampler, while all other parameters are targets for model transition.  We set the scale hyperparameter to $\sigma=12$, $m=\underset{t}{\max}\{y_t\} = 1{,}112$, and the inclusion probability for each parameter to 0.5. All parameters are initialized as 0 in \texttt{Nimble}. The prior distributions are specified as follows: $\beta_0\sim N(0,0.3^2)$, $\beta_1\sim N(0,20^2)$, $\phi_i\sim N(0,0.2^2)$ and $\theta_i\sim N(0,0.2^2)$, for $i\in\{1,2,3\}$. After testing, we determined that a single chain containing 25,000 iterations, with the first 20,000 discarded as burn-in, produced converging chains.

\begin{figure}[!ht]
    \centering
    \includegraphics[width=0.7\textwidth]{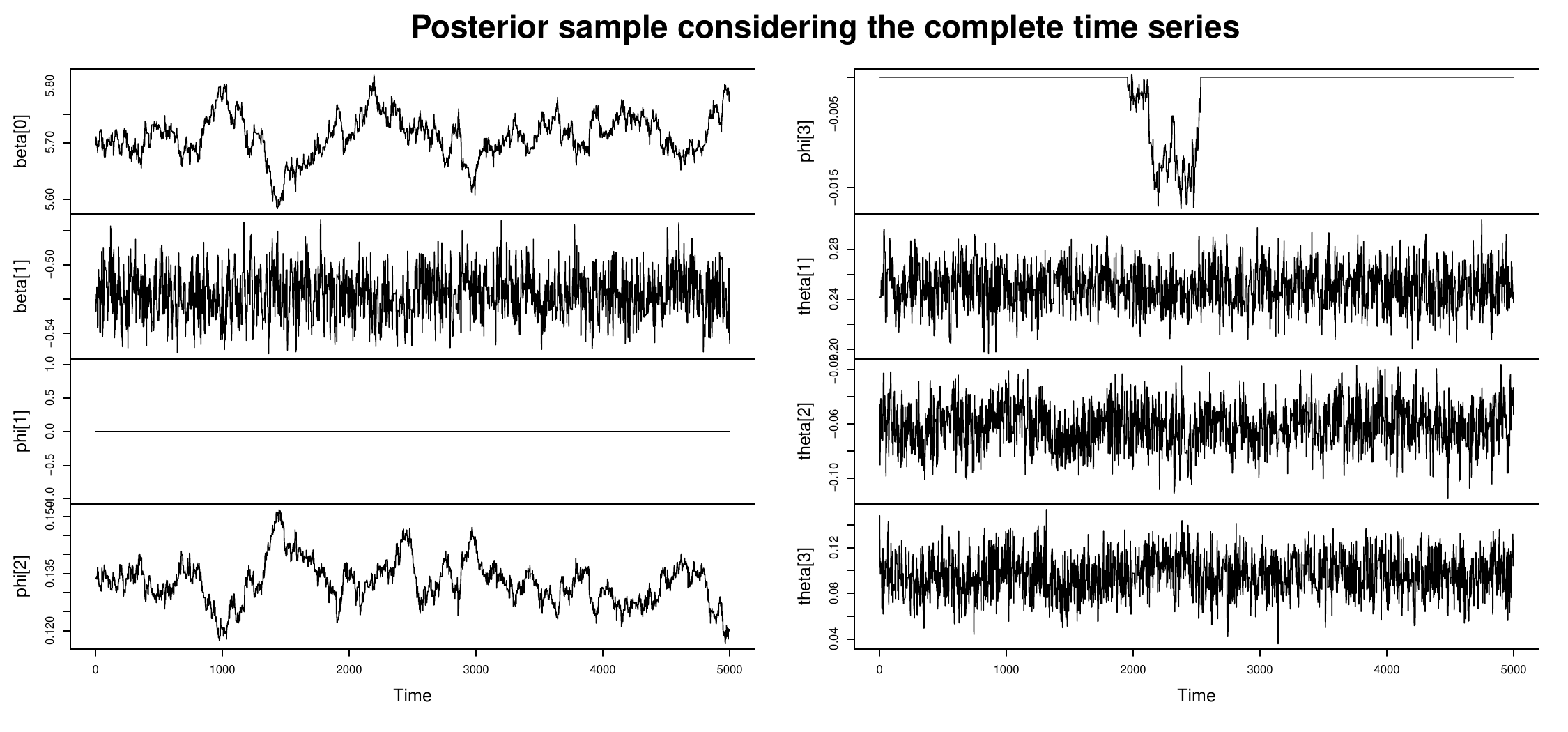}
    \caption{Time series plot of the sample from the posterior distribution for the complete data.}
    \label{tsm2}
\end{figure}

\begin{figure}[!ht]
    \centering
    \includegraphics[width=0.7\textwidth]{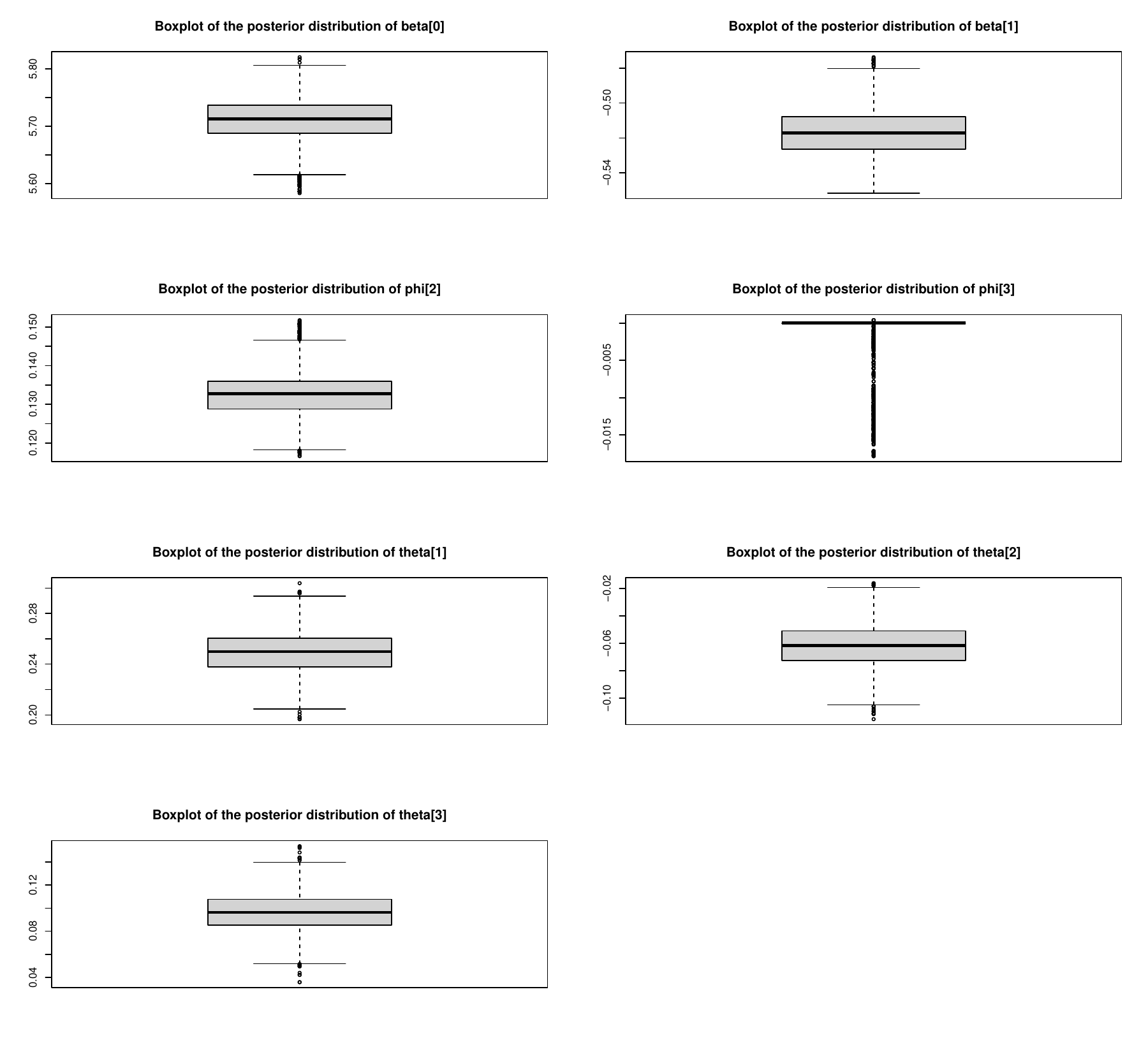}
    \caption{Boxplots of the sample from the posterior distribution for the complete data. For parameter $\phi_1$, the sample from the posterior is constant so that the boxplot was omitted.}
    \label{bpm2}
\end{figure}

Figure \ref{tsm2} present time series plots of the posterior samples obtained for each parameter. Applying the GCD to the samples indicated convergence for all parameters, with the maximum absolute value of the $z$-score obtained being 1.332. Table \ref{buscomp}, presents the mean, median  and $95\%$ HPD credibility interval for each parameter based on the posterior samples. From Table \ref{buscomp},it can be observed that $\phi_1$ and $\phi_3$ are non-significant, pointing to a NB-GARMA$(2,3)$ model with $\phi_1=0$ and $\phi_3=0$. Figure  \ref{bpm2} presents boxplots for the posterior samples for each parameter, (except for $\phi_1$). Observe that the plots are fairly symmetrical in most cases, explaining the close or equal mean and median estimates observed in most cases. Additionally, there is small variability in the samples.

\begin{table}
\scriptsize
\centering
\renewcommand{\arraystretch}{1.4}
\setlength{\tabcolsep}{2pt}
\caption{Summary results obtained from the posterior distribution considering the RJMCMC NB-GARMA approach for the complete time series. Presented are the mean (left) and median (right) along with the HDP credibility interval (below), for each model parameter.}\vspace{.3cm}\label{buscomp}
\begin{tabular}{cc|cc|cc|cc|cc|cc|cc|cc}
\hline
 \multicolumn{2}{c|}{$\beta_0$} & \multicolumn{2}{c|}{$\beta_1$} & \multicolumn{2}{c|}{$\phi_1$}  & \multicolumn{2}{c|}{$\phi_2$}  & \multicolumn{2}{c|}{$\phi_3$} &  \multicolumn{2}{c|}{$\theta_1$} & \multicolumn{2}{c|}{$\theta_2$} & \multicolumn{2}{c}{$\theta_3$} \\
\hline
 5.711  &  5.713  &  -0.517  &  -0.517  &  0.000  &  0.000  &  0.133  &  0.133  &  -0.001  &  0.000  &  0.250  &  0.250  &  -0.062  &  -0.062  &  0.096  &  0.096  \\
  \multicolumn{2}{c|}{$[5.650, 5.800]$}  &   \multicolumn{2}{c|}{$[-0.541, -0.492]$}  &   \multicolumn{2}{c|}{$[0.000, 0.000]$}  &   \multicolumn{2}{c|}{$[0.122, 0.146]$}  &   \multicolumn{2}{c|}{$[-0.011, 0.000]$}  &   \multicolumn{2}{c|}{$[0.220, 0.279]$}  &   \multicolumn{2}{c|}{$[-0.092, -0.033]$}  &   \multicolumn{2}{c}{$[0.065, 0.128]$}  \\
\hline
\end{tabular}
\end{table}
\FloatBarrier

\FloatBarrier
Next, we partitioned the data into two subgroups: the first subgroup consists of $y_1,\cdots,y_{61}$ (sample size 61), representing the period before the structural break, and and the second subgroup consists of $y_{62},\cdots,y_{111}$ (sample size 50), representing the period after the structural break, respectively. For each subgroup, we applied the proposed RJMCMC approach to fit a NB-GARMA model considering \eqref{app2} without the covariate.
%Due to the smaller sample size, calibrating the chain to achieve convergence was more challenging.
For the subgroup before the pandemic, the RJMCMC setup was the same as that used for the complete data, except that the covariate was excluded from the model. {Similarly, for the subgroup after the pandemic, the setup was the same. In both cases, a shorter chain was sufficient to achieve convergence. After experimentation, we found that for data before the change point, a single chain of size 3,000 with the first 2,000 observation discarded as burn-in produced converging chains. For data after the pandemics, a chain of size 6,000 with the first 4,000 observations discarded as burn-in was found to be sufficient. In both cases, the samples from the posterior distribution were tested and found to be convergent using GCD,} with a $z$-score of 1.96 used as the threshold for convergence.

\begin{figure}[!ht]
    \centering
    \includegraphics[width=0.7\textwidth]{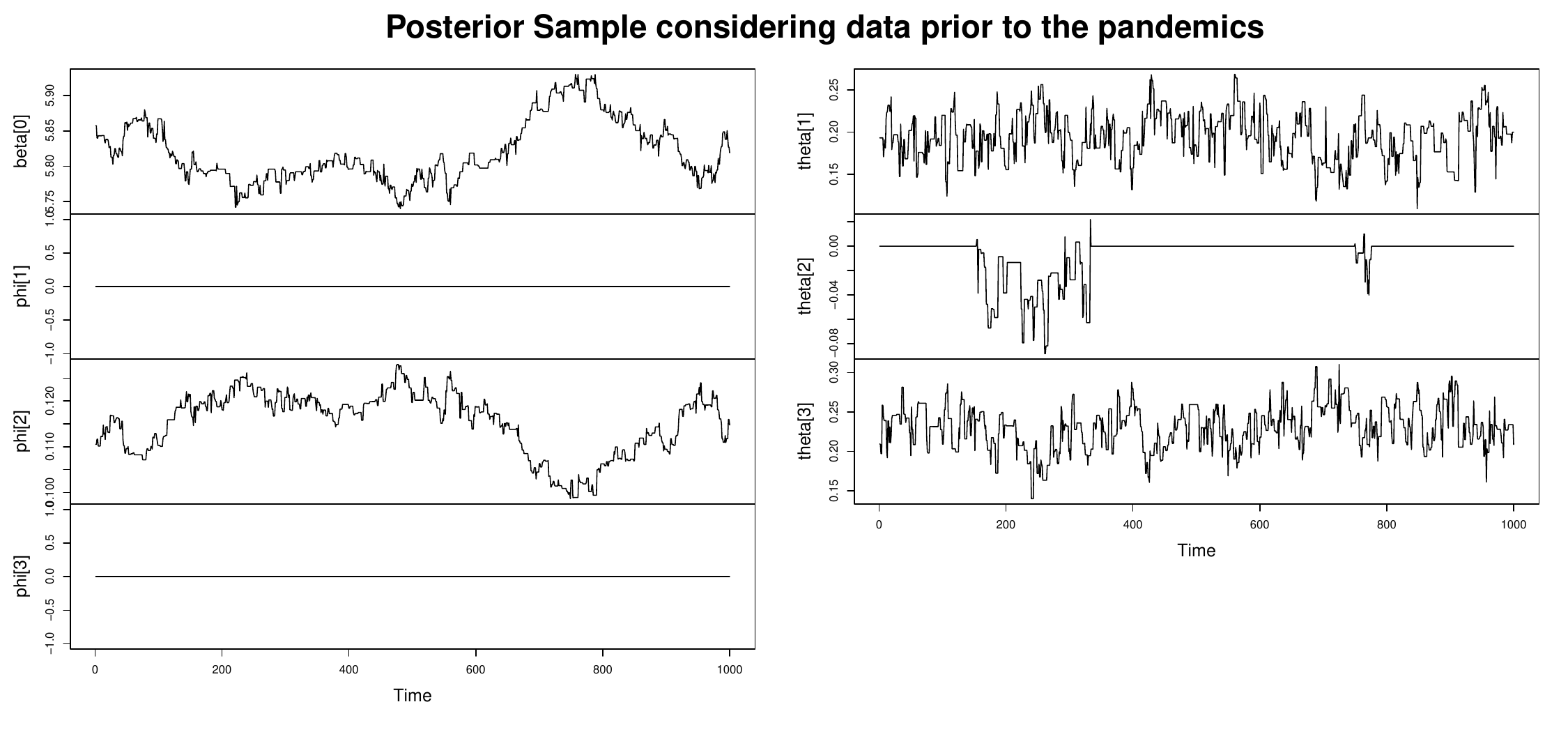}
    \caption{Time series plot of the sample from the posterior distribution before the pandemic.}
    \label{tsmb2}
\end{figure}

\begin{figure}[!ht]
    \centering
    \includegraphics[width=0.7\textwidth]{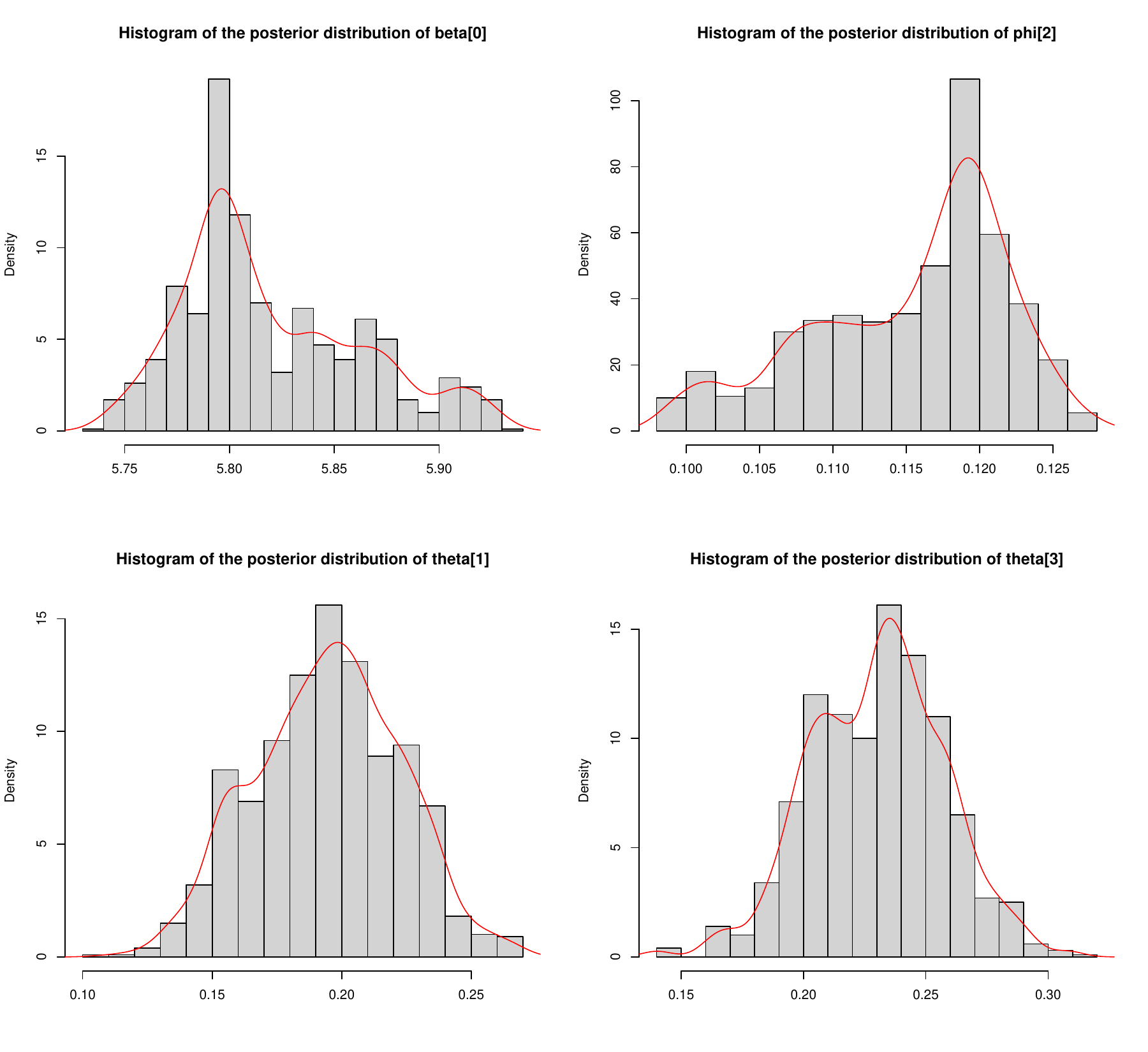}
    \caption{Histograms along with the kernel density estimation of the sample from the posterior distribution before the pandemic. For parameters $\phi_1$, $\phi_3$, and $\theta_2$ the sample from the posterior was almost constant so that the histograms were omitted.}
    \label{hmb2}
\end{figure}

\begin{figure}[!ht]
    \centering
    \includegraphics[width=0.7\textwidth]{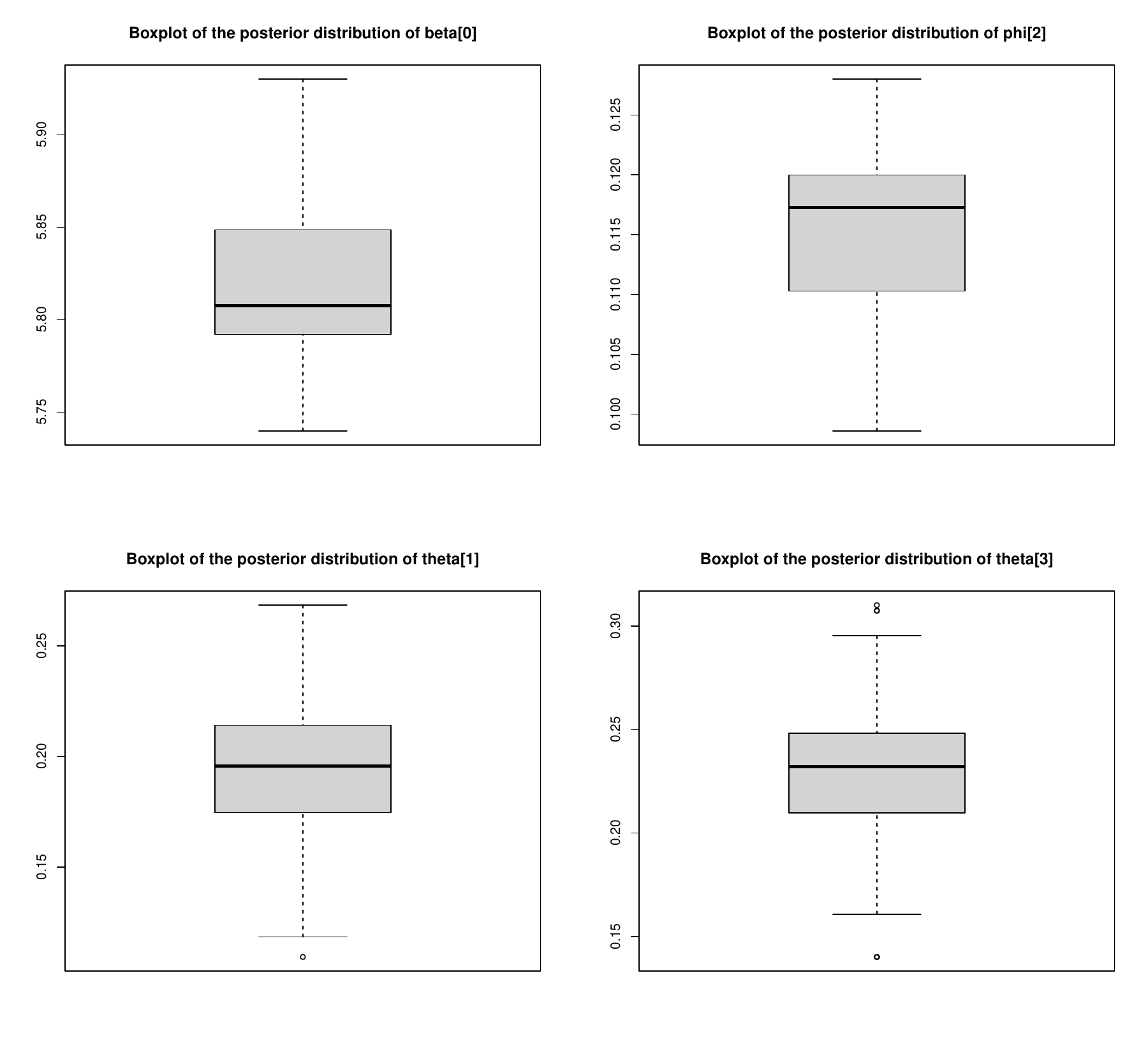}
    \caption{Boxplots of the sample from the posterior distribution before the pandemic. For parameters $\phi_1$, $\phi_3$ and $\theta_2$, the sample from the posterior was almost constant so that the boxplots were omitted.}
    \label{bpmb2}
\end{figure}

\begin{figure}[!ht]
    \centering
    \includegraphics[width=\textwidth]{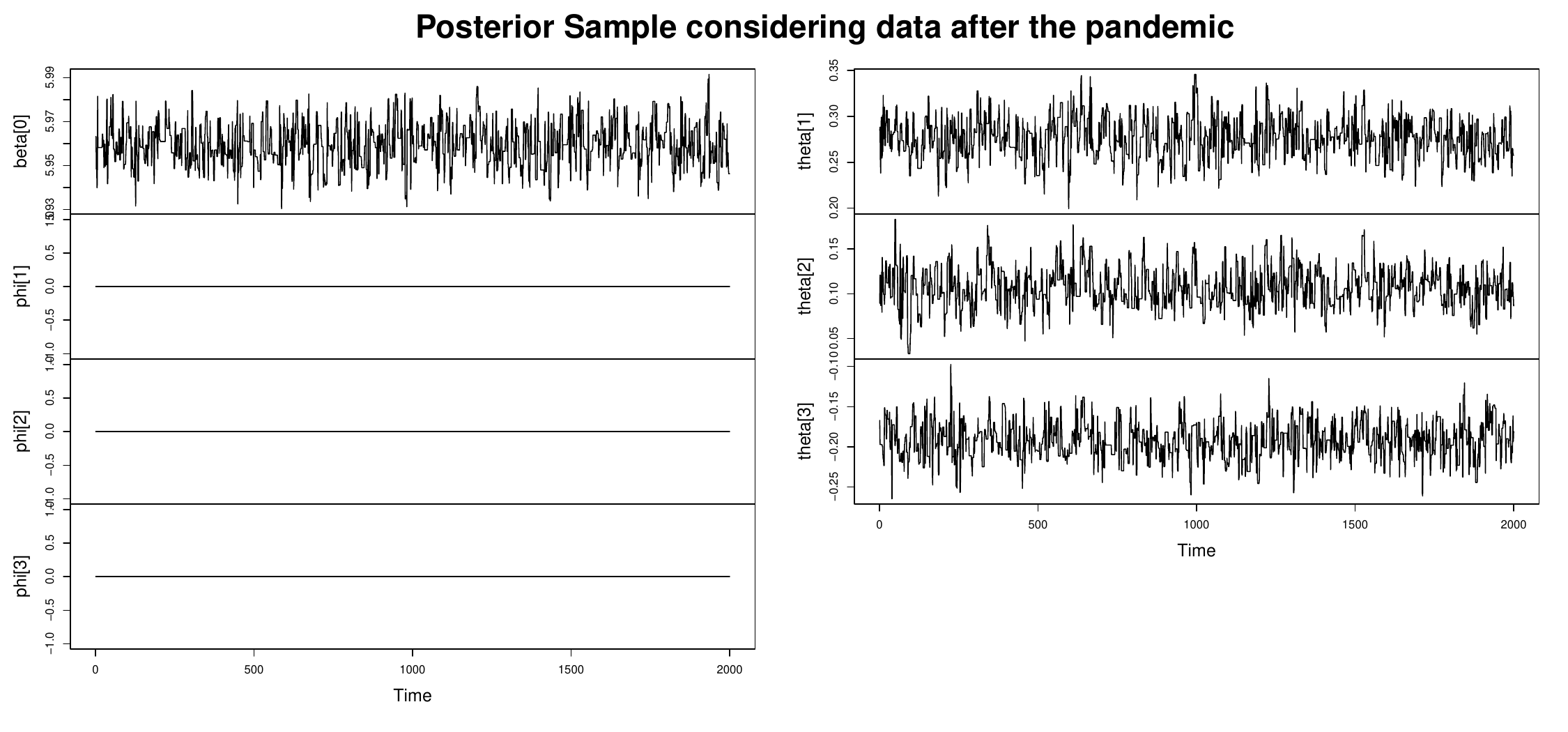}
    \caption{Time series plot of the sample from the posterior distribution after the pandemic.}
    \label{tsma2}
\end{figure}

\begin{figure}[!ht]
    \centering
    \includegraphics[width=0.7\textwidth]{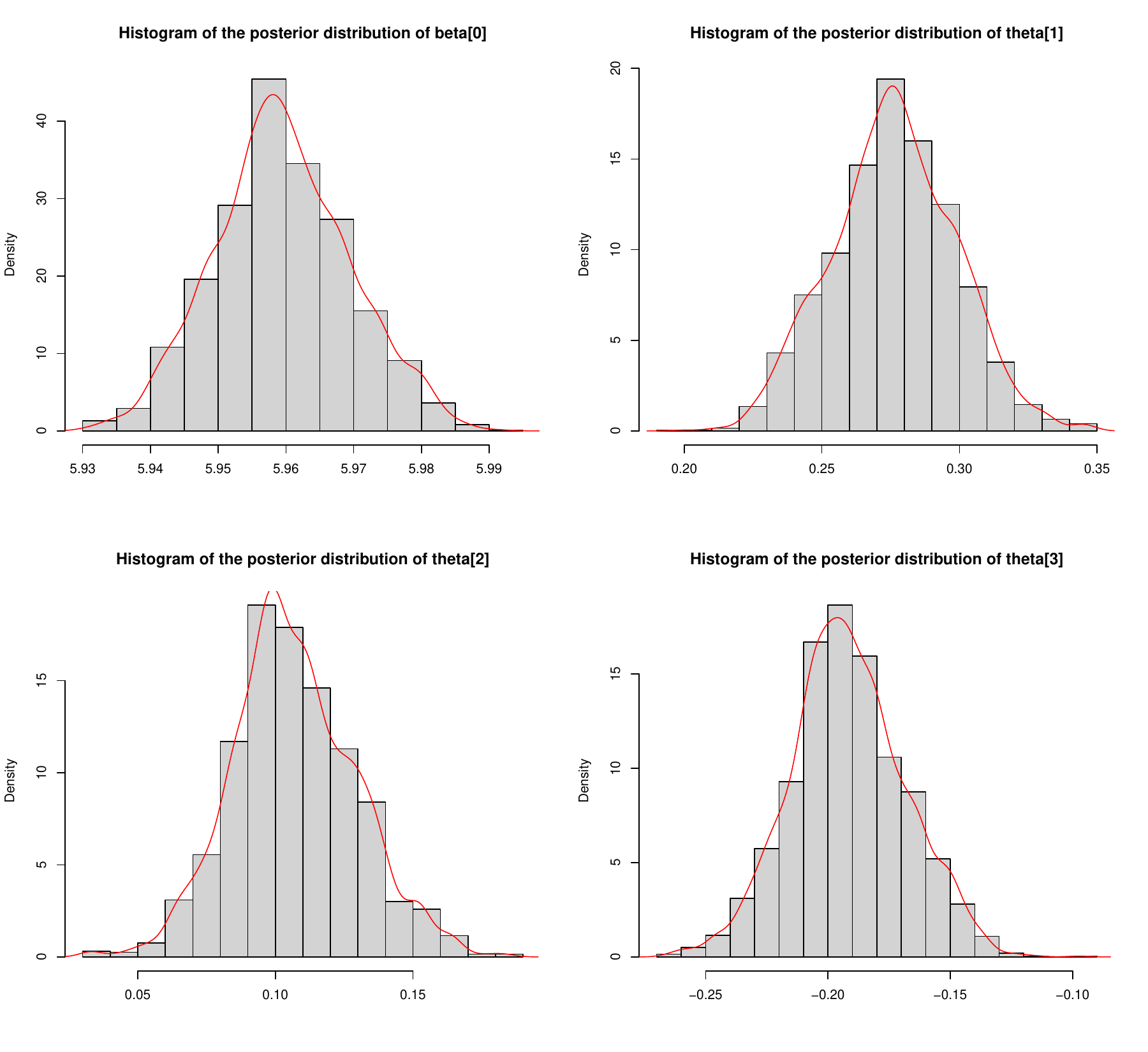}
    \caption{Histograms along with the kernel density estimation of the sample from the posterior distribution after the pandemic. For parameters $\phi_1$, $\phi_2$, and $\phi_3$, the sample from the posterior were constant so that the histograms were omitted.}
    \label{hma2}
\end{figure}

\begin{figure}[!ht]
    \centering
    \includegraphics[width=0.7\textwidth]{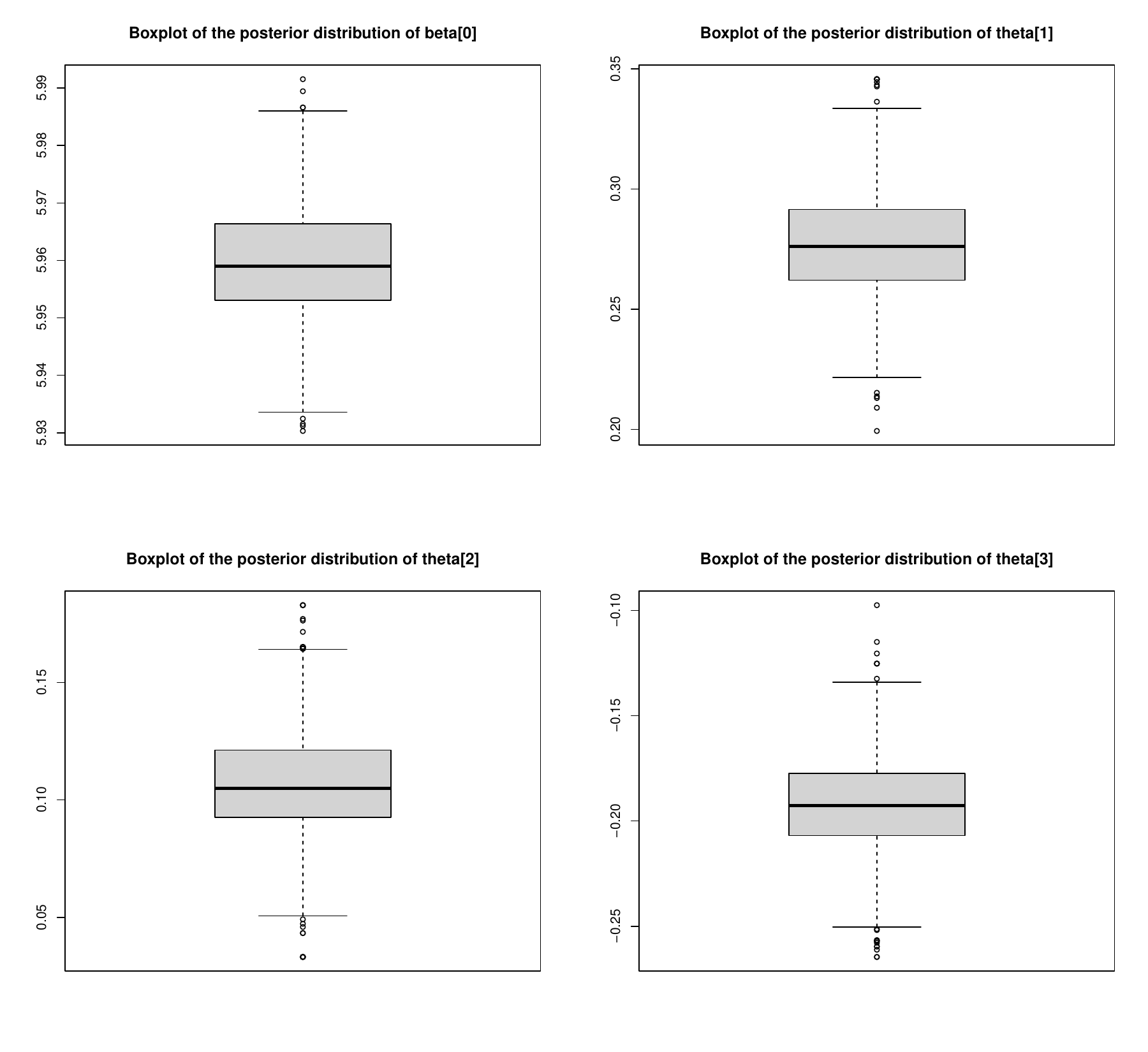}
    \caption{Boxplots of the sample from the posterior distribution after the pandemic. For parameters $\phi_1$, $\phi_2$, and $\phi_3$, the sample from the posterior were constant, so that the boxplots were omitted.}
    \label{bpma2}
\end{figure}

A summary of the results is presented in Table \ref{busparts}. Time series plot, histograms, and boxplots of the posterior sample for the data before the structural break are presented in Figures \ref{tsmb2}, \ref{hmb2}, and \ref{bpmb2}, respectively. Similarly, for the data after the structural break, the plots are shown in Figures  \ref{tsma2}, \ref{hma2}, and \ref{bpma2}. Overall, we observe that the posterior sample is fairly symmetrical for all parameter resulting in similar values for the mean and median in both scenarios. Variability is also small in all cases.

\begin{figure}[!ht]
    \centering
    \includegraphics[width=0.6\textwidth]{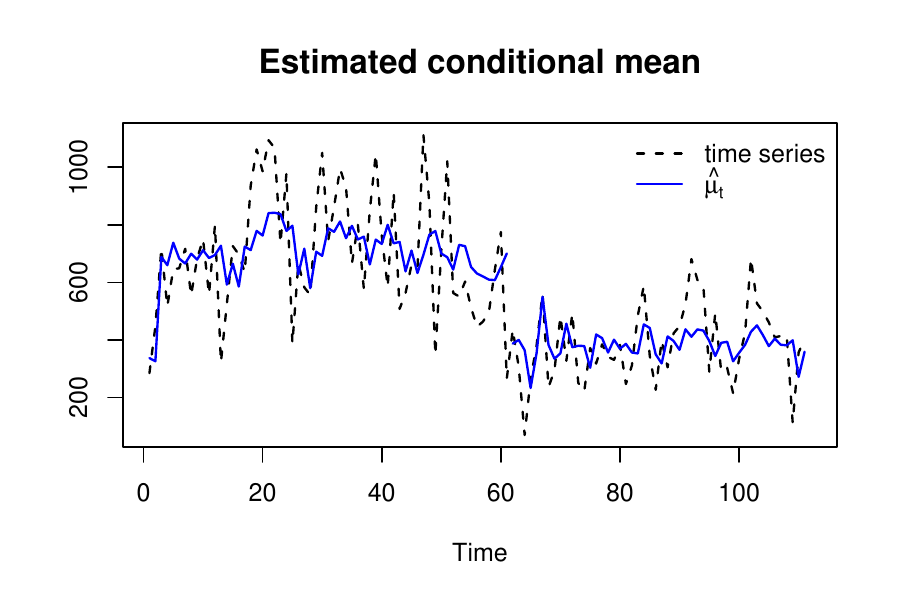}
    \caption{The plot shows the time series along with  $\hat\mu_t$ obtained from the fitted model before and after the pandemic, considering the posterior mean as point estimate.}
    \label{overallfit}
    \end{figure}

Based on the 95\% HPD credible intervals, the model selected for the data before the structural break is a NB-GARMA$(2,3)$ with $\phi_1=0$ and $\theta_2=0$, whereas for the data after the structural break, a full NB-GARMA$(0,3)$ was selected. {It is noteworthy that the estimated value of $\hat\beta_0$ is higher after the pandemic than before, despite the average production being greater in the latter period. This arises because the dynamics of the conditional mean are primarily driven by the time series component. The absence of the AR term in the post-pandemic model results in a higher $\hat\beta_0$ compared to the pre-pandemic model. These findings suggest that the pandemic prompted a shift from a model where the number of buses exported two months ago significantly influenced current exports to one where this dynamic effect has disappeared. Figure \ref{overallfit} we present the reconstructed conditional mean $\mu_t$ based on the (mean) estimated values along with the original time series.}
%It is curious that the estimated value of $\beta_0$ is higher after the pandemics than before even though the average production is higher for the latter. This is a consequence from the fact that the dynamics in the conditional mean is dominated by the time series component. The absence of the AR part in the model after the pandemics causes a higher value of $\beta_0$ compared to the model before the pandemic. These results suggest that the pandemic led to a transition from a model in which the number of buses exported two months ago influenced the exportation of buses this month to a model in which this dynamic has vanished.
%

\begin{table}[ht!]
\scriptsize
\centering
\renewcommand{\arraystretch}{1.3}
\setlength{\tabcolsep}{4pt}
\caption{Summary results from the posterior distribution obtained considering the RJMCMC NB-GARMA approach before and after the structural change. In each cell are presented  the mean (left) and median (right) along with the HDP credibility interval (below) for each parameter.}\vspace{.3cm}\label{busparts}
\begin{tabular}{cc|cc|cc|cc|cc|cc|cc}
\multicolumn{14}{c}{Before}\\
\hline
\multicolumn{2}{c|}{$\beta_0$} &  \multicolumn{2}{c|}{$\phi_1$}  & \multicolumn{2}{c|}{$\phi_2$}  & \multicolumn{2}{c|}{$\phi_3$} &  \multicolumn{2}{c|}{$\theta_1$} & \multicolumn{2}{c|}{$\theta_2$} & \multicolumn{2}{c}{$\theta_3$} \\
\hline
 5.820  &  5.808  &  0.000  &  0.000  &  0.115  &  0.117  &  0.000  &  0.000  &  0.194  &  0.196  &  -0.006  &  0.000  &  0.230  &  0.232  \\
\multicolumn{2}{c|}{$[ 5.755 ,  5.915 ]$} & \multicolumn{2}{c|}{$[ 0.000 ,  0.000 ]$} & \multicolumn{2}{c|}{$[ 0.101 ,  0.125 ]$} & \multicolumn{2}{c|}{$[ 0.000 ,  0.000 ]$} & \multicolumn{2}{c|}{$[ 0.140 ,  0.244 ]$} & \multicolumn{2}{c|}{$[ -0.052 ,  0.000 ]$} & \multicolumn{2}{c}{$[ 0.182 ,  0.285 ]$} \\
\hline\hline
 \multicolumn{14}{c}{After}\\
  \hline
\multicolumn{2}{c|}{$\beta_0$} &  \multicolumn{2}{c|}{$\phi_1$}  & \multicolumn{2}{c|}{$\phi_2$}  & \multicolumn{2}{c|}{$\phi_3$} &  \multicolumn{2}{c|}{$\theta_1$} & \multicolumn{2}{c|}{$\theta_2$} & \multicolumn{2}{c}{$\theta_3$} \\
\hline
5.960  &  5.959  &  0.000  &  0.000  &  0.000  &  0.000  &  0.000  &  0.000  &  0.276  &  0.276  &  0.107  &  0.105  &  -0.192  &  -0.193  \\
\multicolumn{2}{c|}{$[ 5.941 ,  5.979 ]$} & \multicolumn{2}{c|}{$[ 0.000 ,  0.000 ]$} & \multicolumn{2}{c|}{$[ 0.000 ,  0.000 ]$} & \multicolumn{2}{c|}{$[ 0.000 ,  0.000 ]$} & \multicolumn{2}{c|}{$[ 0.231 ,  0.317 ]$} & \multicolumn{2}{c|}{$[ 0.065 ,  0.153 ]$} & \multicolumn{2}{c}{$[ -0.233 ,  -0.143 ]$} \\
\hline\hline
\end{tabular}
\end{table}
\FloatBarrier
An intriguing finding is that the model considering the complete dataset along with the dummy variable exhibits same order and somewhat comparable coefficients, especially the autoregressive one, with the model before the pandemics. This suggests that the combined model is predominantly influenced by the data before the pandemic, which comprises 22\% more observations than the data after the pandemic. Consequently, the model averages out both dynamics, effectively tying them together through the dummy variable.

It is important to note that this analysis has limitations, as we did not consider other external factors that could explain these changes, such as logistical limitations imposed by the pandemic, changes in commercial arrangements, or external economic factors. However, the primary objective was to explore the potential of the proposed methodology in this context rather than engage in a comprehensive economic discussion of this significant topic.

\FloatBarrier
\section{Conclusion}
In this paper, we tackle the problem of order selection in GARMA models for count time series from a Bayesian perspective, using the approach known as \emph{Reversible Jump Markov Chain Monte Carlo} (RJMCMC). The study successfully achieved its main objective of investigating the selection of GARMA count models in the Bayesian context, through the RJMCMC approach. The sensitivity analysis regarding the choice of hyperparameters for the priors was also addressed, providing valuable insights into the method's robustness and flexibility. The RJMCMC simulations revealed that the implementation of a burn-in  is consistently beneficial, resulting in notable improvements in all cases and metrics. This effect is particularly evident in the significant reduction of the impact of $\sigma$, making the results more reliable, with a notable improvement in the correct identification of models. Applying a burn-in showed significant improvements for GAR$(1)$, while for GAR$(2)$ the benefits were less pronounced, and for the GMA$(q)$ model, the influence when applying a burn-in  on the point estimate is significantly less notable compared to the GAR$(p)$ model.

In contrast, the application of thinning between lags did not produce substantial improvements in point estimation or effective sample size, indicating that application of this procedure in the context of GARMA models is not advisable. In section 4, we address the empirical application in real-world datasets, which demonstrated the practical relevance of the proposed method, highlighting its ability to handle real-world situations.
\subsection*{Acknowledgments}
Katerine Zuniga Lastra gratefully acknowledges the financial support granted by the Coordena\c c\~ao de Aperfei\c coamento de Pessoal de N\'ivel Superior – Brazil (CAPES) -- Programa CAPES-DS.
\subsection*{Statements and Declarations}
The authors declare that they have NO affiliations with or involvement in any organization or entity with any financial interests in the subject matter or materials discussed in this manuscript.

\FloatBarrier
\bibliographystyle{apalike}
\bibliography{kate}

\appendix

\section{Tables of section \ref{BI} - the effects of burn-in }

\begin{table}
\scriptsize
\centering
\renewcommand{\arraystretch}{1.1}
\setlength{\tabcolsep}{4pt}
%\caption{Tabela burn-in M1: applying a burning improves the accuracy of the estimates, eliminating the effects of choosing the scale to run the chains.}\vspace{.3cm}
\caption{Simulation Results for GAR$(1)$ Models with burn-in $\{0, 1000, 3000, 5000\}$ and $\sigma\in\{0.5,5,10,15\}$.}\vspace{.3cm}\label{burn-inGAR1}
\begin{tabular}{c|c|c|c|cccc|c|c|cccc}
\hline
 \multirow{2}{*}{Burn-in}&Param-   &  \multicolumn{6}{c|}{$\sigma = 0.5$} &  \multicolumn{6}{c}{$\sigma = 5$} \\
 \cline{3-14}
&eter & HPD & ECI& Mean & Med & SD & ESS & HPD & ECI& Mean & Med & SD & ESS \\
 \hline
\multirow{7}{*}{0} & $\alpha=-0.5$ & \multirow{7}{*}{74.0\%} & \multirow{7}{*}{66.4\%} & -0.509 & -0.503 & 0.149 & 70 & \multirow{7}{*}{55.8\%} & \multirow{7}{*}{48.2\%} & -0.524 & -0.508 & 0.160 & 54 \\
   & $\phi_1=-0.4$  &   &   & -0.369 & -0.385 & 0.093 & 81 &   &   & -0.358 & -0.382 & 0.104 & 55 \\
   & $\phi_2=0.0$   &   &   & -0.012 & -0.001 & 0.054 & 114 &   &   & -0.010 & -0.001 & 0.049 & 88 \\
   & $\phi_3=0.0$   &   &   & -0.011 & -0.003 & 0.043 & 185 &   &   & -0.012 & -0.003 & 0.042 & 133 \\
   & $\theta_1=0.0$ &   &   & -0.027 & -0.002 & 0.090 & 71 &   &   & -0.037 & -0.003 & 0.100 & 45 \\
   & $\theta_2=0.0$ &   &   & 0.018 & 0.002 & 0.061 & 147 &   &   & 0.019 & 0.003 & 0.058 & 121 \\
\hline
 \end{tabular}
\end{table}

\FloatBarrier
\begin{table}
\scriptsize
\centering
\renewcommand{\arraystretch}{1.1}
\setlength{\tabcolsep}{4pt}
%\caption{Tabela burn-in M2: applying a burning improves the accuracy of the estimates, eliminating the effects of choosing the scale to run the chains.}\vspace{.3cm}
\caption{Simulation Results for GAR$(2)$ Models considering burn-in $\{0,1000, 3000, 5000\}$ and $\sigma\in\{0.5,5,10,15\}$.}\vspace{.3cm}\label{burn-inGAR2}
\begin{tabular}{c|c|c|c|cccc|c|c|cccc}
\hline
\multirow{2}{*}{Burn-in}&Param-  &  \multicolumn{6}{c|}{$\sigma = 0.5$} &  \multicolumn{6}{c}{$\sigma = 5$} \\
\cline{3-14}
&eter & HPD & ECI& Mean & Med & SD & ESS & HPD & ECI& Mean & Med & SD & ESS \\
    \hline
 \multirow{7}{*}{0} & $\alpha=-1.0$ & \multirow{7}{*}{98.7\%} & \multirow{7}{*}{97.3\%} & -0.942 & -0.966
 & 0.135 & 85 & \multirow{7}{*}{95.9\%} & \multirow{7}{*}{89.0\%} & -0.949 & -0.969 & 0.127 & 76 \\
   & $\phi_1=0.0$  &   &   & -0.036 & -0.006 & 0.082 & 77 &   &   & -0.033 & -0.006 & 0.074 & 70 \\
   & $\phi_2=-0.4$   &   &   & -0.394 & -0.396 & 0.063 & 234 &   &   & -0.391 & -0.395 & 0.068 & 164 \\
   & $\phi_3=0.0$   &   &   & -0.021 & -0.006 & 0.049 & 199 &   &   & -0.019 & -0.006 & 0.045 & 153 \\
   & $\theta_1=0.0$ &   &   & 0.034 & 0.005 & 0.083 & 83 &   &   & 0.031 & 0.004 & 0.076 & 83 \\
   & $\theta_2=0.0$ &   &   & -0.002 & -0.000 & 0.062 & 210 &   &   & -0.004 & -0.000 & 0.067 & 135 \\
\hline
\end{tabular}
\end{table}

\FloatBarrier
\begin{table}
\scriptsize
\centering
\renewcommand{\arraystretch}{1.1}
\setlength{\tabcolsep}{4pt}
%\caption{Tabela burn-in M3: applying a burning improves the accuracy of the estimates, eliminating the effects of choosing the scale to run the chains.}\vspace{.3cm}
\caption{RJMCMC Simulation Results for GMA$(1)$ Models considering burn-in $\{0,1000, 3000, 5000\}$ and $\sigma\in\{0.5,5,10,15\}$.}\vspace{.3cm}\label{burn-inGMA1}
\begin{tabular}{c|c|c|c|cccc|c|c|cccc}
\hline
 \multirow{2}{*}{Burn-in}&Param-  &  \multicolumn{6}{c|}{$\sigma = 0.5$} &  \multicolumn{6}{c}{$\sigma = 5$} \\
 \cline{3-14}
 &eter  & HPD & ECI& Mean & Med & SD & ESS & HPD & ECI& Mean & Med & SD & ESS \\
 \hline
\multirow{7}{*}{0} & $\alpha=-0.5$ & \multirow{7}{*}{94.0\%} & \multirow{7}{*}{93.7\%} & -0.356 & -0.389 & 0.153 & 11&
\multirow{7}{*}{89.2\%} & \multirow{7}{*}{88.4\%} & -0.277 & -0.301 & 0.181 & 11  \\

   & $\phi_1=0.0$  &    &   & -0.044 & -0.028 & 0.050 & 31   &   &  & -0.075 & -0.061 & 0.066 & 14     \\
   & $\phi_2=0.0$   &   &   & -0.005 & -0.002 & 0.025 & 36   &   &  & -0.006 & -0.002 & 0.033 & 28     \\
   & $\phi_3=0.0$   &   &   & -0.003 & -0.002 & 0.022 & 43   &   &  & -0.001 & 0.001  & 0.029 & 35     \\
   & $\theta_1=-0.5$ &  &   & -0.430 & -0.437 & 0.071 & 1037 &   &  & -0.402 & -0.408 & 0.080 & 451    \\
   & $\theta_2=0.0$ &   &   & -0.005 & -0.002 & 0.041 & 1305 &   &  & -0.011 & -0.005 & 0.049 & 1288   \\
\hline
\end{tabular}
\end{table}

\FloatBarrier
\begin{table}
\scriptsize
\centering
\renewcommand{\arraystretch}{1.1}
\setlength{\tabcolsep}{4pt}
%\caption{Tabela burn-in M4: applying a burning improves the accuracy of the estimates, eliminating the effects of choosing the scale to run the chains.}\vspace{.3cm}
\caption{RJMCMC Simulation Results for GMA$(2)$ Models considering burn-in $\{0,1000, 3000, 5000\}$ and $\sigma\in\{0.5,5,10,15\}$.}\vspace{.3cm}\label{burn-inGMA2}
\begin{tabular}{c|c|c|c|cccc|c|c|cccc}
\hline
\multirow{2}{*}{Burn-in}&Param-  &  \multicolumn{6}{c|}{$\sigma = 0.5$} &  \multicolumn{6}{c}{$\sigma = 5$} \\
 \cline{3-14}
&eter & HPD & ECI& Mean & Med & SD & ESS & HPD & ECI& Mean & Med & SD & ESS \\
 \hline
\multirow{7}{*}{0} & $\alpha=-1$ & \multirow{7}{*}{{99.4\%}} & \multirow{7}{*}{{99.4\%}} & -0.929 & -0.951 & 0.121 & 125
& \multirow{7}{*}{98.9\%} & \multirow{7}{*}{99.1\%} & -0.813 & -0.843 & 0.191 & 32 \\
   & $\phi_1=0.0$   &   &   & -0.029 & -0.007 & 0.067 & 102&   &   & -0.050 & -0.027 & 0.072 & 33   \\
   & $\phi_2=0.0$   &   &   & -0.005 & -0.001 & 0.048 & 212&   &   & -0.009 & -0.003 & 0.047 & 48   \\
   & $\phi_3=0.0$   &   &   & -0.014 & -0.006 & 0.038 & 281&   &   & -0.020 & -0.010 & 0.044 & 61   \\
   & $\theta_1=0.0$ &   &   & 0.025  & 0.003  & 0.065 & 116&   &   & 0.040  & 0.017  & 0.071 & 127  \\
   & $\theta_2=0.6$ &   &   & 0.587  & 0.588  & 0.052 & 518&   &   & 0.586  & 0.586  & 0.053 & 290  \\
\hline
\end{tabular}
\end{table}
\FloatBarrier

\section{Tables of Section \ref{ET} - the effects of thinning}
\FloatBarrier
\begin{table}
\scriptsize
\centering
\renewcommand{\arraystretch}{1.1}
\setlength{\tabcolsep}{4pt}
%\caption{Tabela thinning M1: applying a burning improves the accuracy of the estimates, eliminating the effects of choosing the scale to run the chains.}\vspace{.3cm}
\caption{RJMCMC Simulation Results for GAR$(1)$ Models considering thinning $\{5,10, 20\}$ and $\sigma\in\{0.5,5,10,15\}$.}\vspace{.3cm}\label{thinningGAR1}
\begin{tabular}{c|c|c|c|cccc|c|c|cccc}
\hline
 \multirow{2}{*}{Thinning}&Param-  &  \multicolumn{6}{c|}{$\sigma = 0.5$} &  \multicolumn{6}{c}{$\sigma = 5$} \\
 \cline{3-14}
& eter & HPD & ECI& Mean & Med & SD & ESS & HPD & ECI& Mean & Med & SD & ESS \\
 \hline
\multirow{7}{*}{5} & $\alpha=-0.5$ & \multirow{7}{*}{{74.0\%}} & \multirow{7}{*}{{66.5\%}} & -0.509 & -0.503 & 0.149 & 63 & \multirow{7}{*}{55.7\%} & \multirow{7}{*}{48.2\%} & -0.524 & -0.508 & 0.160 & 47 \\
   & $\phi_1=-0.4$  &   &   & -0.369 & -0.385 & 0.093 & 68 &   &   & -0.358 & -0.382 & 0.104 & 45 \\
   & $\phi_2=0.0$   &   &   & -0.012 & -0.001 & 0.054 & 103 &   &   & -0.010 & -0.001 & 0.049 & 83 \\
   & $\phi_3=0.0$   &   &   & -0.011 & -0.003 & 0.043 & 166 &   &   & -0.012 & -0.003 & 0.042 & 125 \\
   & $\theta_1=0.0$ &   &   & -0.027 & -0.002 & 0.090 & 65 &   &   & -0.037 & -0.003 & 0.100 & 42 \\
   & $\theta_2=0.0$ &   &   & 0.018 & 0.002 & 0.061 & 128 &   &   & 0.019 & 0.003 & 0.058 & 99 \\
\hline
\end{tabular}
\end{table}

\FloatBarrier
\begin{table}
\scriptsize
\centering
\renewcommand{\arraystretch}{1.1}
\setlength{\tabcolsep}{4pt}
\caption{RJMCMC Simulation Results for GAR$(2)$ Models considering thinning $\{5,10, 20\}$ and $\sigma\in\{0.5,5,10,15\}$.}\vspace{.3cm}\label{thinningGAR2}
\begin{tabular}{c|c|c|c|cccc|c|c|cccc}
\hline
 \multirow{2}{*}{Thinning}& param-  &  \multicolumn{6}{c|}{$\sigma = 0.5$} &  \multicolumn{6}{c}{$\sigma = 5$} \\
 \cline{3-14}
& eter & HPD & ECI& Mean & Med & SD & ESS & HPD & ECI& Mean & Med & SD & ESS \\
 \hline
\multirow{7}{*}{5} & $\alpha=-1$ & \multirow{7}{*}{98.7\%} & \multirow{7}{*}{97.2\%} & -0.942 & -0.966 & 0.136 & 74 & \multirow{7}{*}{95.9\%} & \multirow{7}{*}{89.2\%} & -0.949 & -0.969 & 0.128 & 68 \\
   & $\phi_1=0.0$  &   &   & -0.036 & -0.006 & 0.082 & 70 &   &   & -0.033 & -0.006 & 0.075 & 65 \\
   & $\phi_2=-0.4$   &   &   & -0.394 & -0.396 & 0.063 & 210 &   &   & -0.391 & -0.395 & 0.069 & 146 \\
   & $\phi_3=0.0$   &   &   & -0.021 & -0.006 & 0.049 & 172 &   &   & -0.019 & -0.006 & 0.045 & 140 \\
   & $\theta_1=0.0$ &   &   & 0.034 & 0.005 & 0.083 & 76 &   &   & 0.031 & 0.004 & 0.076 & 75 \\
   & $\theta_2=0.0$ &   &   & -0.002 & -0.000 & 0.062 & 194 &   &   & -0.004 & -0.000 & 0.067 & 127 \\
\hline
\end{tabular}
\end{table}
\FloatBarrier
\begin{table}
\scriptsize
\centering
\renewcommand{\arraystretch}{1.1}
\setlength{\tabcolsep}{4pt}
\caption{Simulation Results for the GMA$(1)$ model considering thinning $\{5,10, 20\}$ and $\sigma\in\{0.5,5,10,15\}$.}\vspace{.3cm}\label{thinningGMA1}
\begin{tabular}{c|c|c|c|cccc|c|c|cccc}
\hline
 \multirow{2}{*}{Thinning}&Param  &  \multicolumn{6}{c|}{$\sigma = 0.5$} &  \multicolumn{6}{c}{$\sigma = 5$} \\
 \cline{3-14}
& eter & HPD & ECI& Mean & Med & SD & ESS & HPD & ECI& Mean & Med & SD & ESS \\
 \hline
\multirow{7}{*}{5} & $\alpha=-0.5$ & \multirow{7}{*}{89.3\%} & \multirow{7}{*}{88.4\%} & -0.277 & -0.301 & 0.181 & 11 & \multirow{7}{*}{94.0\%} & \multirow{7}{*}{93.7\%} & -0.356 & -0.389 & 0.153 & 11 \\
   & $\phi_1=0.0$  &   &   & -0.075 & -0.061 & 0.066 & 14 &   &   & -0.044 & -0.028 & 0.050 & 33 \\
   & $\phi_2=0.0$   &   &   & -0.006 & -0.002 & 0.033 & 29 &   &   & -0.005 & -0.002 & 0.025 & 40 \\
   & $\phi_3=0.0$   &   &   & -0.001 & 0.001 & 0.029 & 36 &   &   & -0.003 & -0.002 & 0.022 & 48 \\
   & $\theta_1=-0.5$ &   &   & -0.402 & -0.408 & 0.080 & 202 &   &   & -0.430 & -0.437 & 0.071 & 610 \\
   & $\theta_2=0.0$ &   &   & -0.011 & -0.005 & 0.049 & 412 &   &   & -0.005 & -0.002 & 0.041 & 829 \\
\hline
\end{tabular}
\end{table}

\FloatBarrier
\begin{table}
\scriptsize
\centering
\renewcommand{\arraystretch}{1.1}
\setlength{\tabcolsep}{4pt}
\caption{RJMCMC Simulation Results for GMA$(2)$ Models considering thinning $\{5,10, 20\}$ and $\sigma\in\{0.5,5,10,15\}$.}\vspace{.3cm}\label{thinningGMA2}
\begin{tabular}{c|c|c|c|cccc|c|c|cccc}
\hline
 \multirow{2}{*}{Thinning}&Param-  &  \multicolumn{6}{c|}{$\sigma = 0.5$} &  \multicolumn{6}{c}{$\sigma = 5$} \\
 \cline{3-14}
& eter& HPD & ECI & Mean & Med & SD & ESS & HPD & ECI & Mean & Med & SD & ESS \\
 \hline
\multirow{7}{*}{5} & $\alpha=-1$ & \multirow{7}{*}{98.8\%} & \multirow{7}{*}{99.1\%} & -0.813 & -0.843 & 0.191 & 31 & \multirow{7}{*}{99.4\%} & \multirow{7}{*}{99.4\%} & -0.929 & -0.951 & 0.121 & 110 \\
   & $\phi_1=0.0$   &   &   & -0.050 & -0.027 & 0.072 & 32 &   &   & -0.029 & -0.007 & 0.067 & 94 \\
   & $\phi_2=0.0$   &   &   & -0.009 & -0.003 & 0.047 & 48 &   &   & -0.005 & -0.001 & 0.048 & 206 \\
   & $\phi_3=0.0$   &   &   & -0.020 & -0.010 & 0.044 & 59 &   &   & -0.014 & -0.006 & 0.038 & 266 \\
   & $\theta_1=0.0$ &   &   & 0.040 & 0.017 & 0.072 & 93 &   &   & 0.025 & 0.003 & 0.065 & 108 \\
   & $\theta_2=0.6$ &   &   & 0.586 & 0.586 & 0.054 & 168 &   &   & 0.587 & 0.588 & 0.052 & 455 \\
\hline
\end{tabular}
\end{table}

\end{document}